\documentclass[twocolumn,twocolappendix,numberedappendix,appendixfloats]{openjournal}

\usepackage{amsmath,amssymb,bm,natbib,latexsym,times}

\usepackage[frozencache]{minted}
\usepackage{listings}
\lstdefinestyle{yaml}{
     basicstyle=\ttfamily\color{black}\scriptsize,
     rulecolor=\color{black},
     keywordstyle=\color{blue}\bfseries,
     comment=[l]{\#},
     commentstyle=\itshape\color{black},
     showstringspaces=false
}

\usepackage[T1]{fontenc}
\usepackage{aecompl}

\usepackage{verbatim}
\usepackage{hhline}
\usepackage[dvipsnames]{xcolor}
\usepackage{todonotes}
\usepackage{xspace}
\usepackage{hyperref}
\usepackage{cuted}
\usepackage{eso-pic}

\hypersetup{colorlinks=true,linkcolor=blue,citecolor=blue,filecolor=blue,urlcolor=blue}

\numberwithin{equation}{section}

\DeclareMathSymbol{:}{\mathord}{operators}{"3A}

\newcommand{\eg}{{\sl e.g.}, }

\newcommand{\Gaia}{\textit{Gaia}\xspace}
\newcommand{\Balrog}{\textsc{Balrog}\xspace}
\newcommand{\mdet}{\textsc{Metadetection}\xspace}
\newcommand{\maglim}{\textsc{Maglim++}\xspace}
\newcommand{\redmagic}{\textsc{RedMaGic}\xspace}
\newcommand{\SE}{\textsc{SourceExtractor}\xspace}
\newcommand{\fitvd}{\textsc{Fitvd}\xspace}

\definecolor{orcidlogocol}{HTML}{A6CE39}
\definecolor{purple}{RGB}{128, 0, 128}
\definecolor{kelly}{RGB}{76, 187, 23}

\newcommand{\OrcidIDName}[2]{\href{https://orcid.org/#1}{#2}}

\defcitealias{BalrogY3}{E22}
\defcitealias{Balrog_Y1}{S16}
\defcitealias{DES_DF_Y3}{HC21}

\newcommand*{\vcenteredhbox}[1]{\begingroup
\setbox0=\hbox{#1}\parbox{\wd0}{\box0}\endgroup}

\begin{document}

\AddToShipoutPictureBG*{%
  \AtPageUpperLeft{%
    \hspace*{18.25cm}%
    \raisebox{-9.4\baselineskip}{%
      \makebox[0pt][l]{\textnormal{DES 2024-0867}}
 
}}}%

\AddToShipoutPictureBG*{%
  \AtPageUpperLeft{%
    \hspace*{20.45cm} 
    \raisebox{-10.5\baselineskip}{%
      \makebox[0pt][r]{\textnormal{FERMILAB-PUB-24-0940-T}}
}}}%

\title{Dark Energy Survey Year 6 Results: Synthetic-source Injection Across the Full Survey Using \Balrog}
\shorttitle{Dark Energy Survey Year 6 Balrog}
\shortauthors{Anbajagane, Tabbutt, Beas-Gonzalez \& Yanny}

\email{dhayaa@uchicago.edu}


\author{\OrcidIDName{0000-0003-3312-909X}{D.~Anbajagane} (\vcenteredhbox{\includegraphics[height=1.2\fontcharht\font`\B]{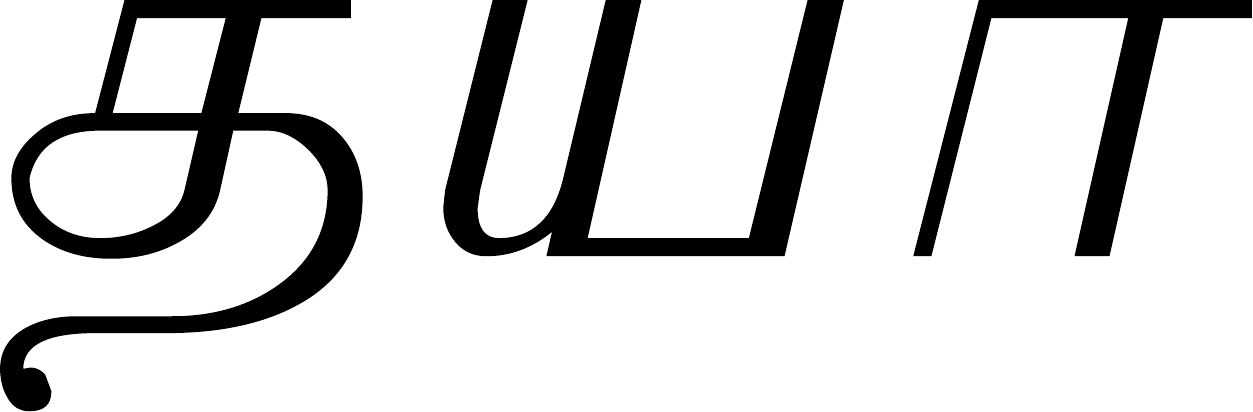}})}
\affiliation{Kavli Institute for Cosmological Physics, University of Chicago, Chicago, IL 60637, USA}
\affiliation{Department of Astronomy and Astrophysics, University of Chicago, Chicago, IL 60637, USA}

\author{\OrcidIDName{0000-0002-0690-1737}{M.~Tabbutt}}
\affiliation{Physics Department, 2320 Chamberlin Hall, University of Wisconsin-Madison, 1150 University Avenue Madison, WI  53706-1390}

\author{\OrcidIDName{0000-0001-8298-7205}{J.~Beas-Gonzalez}}
\affiliation{Physics Department, 2320 Chamberlin Hall, University of Wisconsin-Madison, 1150 University Avenue Madison, WI  53706-1390}

\author{\OrcidIDName{0000-0002-9541-2678}{B.~Yanny}}
\affiliation{Fermi National Accelerator Laboratory, P. O. Box 500, Batavia, IL 60510, USA}

\author{S.~Everett}
\affiliation{California Institute of Technology, 1200 East California Blvd, MC 249-17, Pasadena, CA 91125, USA}

\author{\OrcidIDName{0000-0001-7774-2246}{M.~R.~Becker}}
\affiliation{Argonne National Laboratory, 9700 South Cass Avenue, Lemont, IL 60439, USA}

\author{\OrcidIDName{0000-0003-1585-997X}{M.~Yamamoto}}
\affiliation{Department of Astrophysical Sciences, Princeton University, Peyton Hall, Princeton, NJ 08544, USA}
\affiliation{Department of Physics, Duke University Durham, NC 27708, USA}

\author{E.~Legnani}
\affiliation{Institut de F\'{\i}sica d'Altes Energies (IFAE), The Barcelona Institute of Science and Technology, Campus UAB, 08193 Bellaterra (Barcelona) Spain}

\author{\OrcidIDName{0000-0001-8318-6813}{J.~De~Vicente}}
\affiliation{Centro de Investigaciones Energ\'eticas, Medioambientales y Tecnol\'ogicas (CIEMAT), Madrid, Spain}

\author{K.~Bechtol}
\affiliation{Physics Department, 2320 Chamberlin Hall, University of Wisconsin-Madison, 1150 University Avenue Madison, WI  53706-1390}

\author{\OrcidIDName{0000-0001-5148-9203}{J.~Elvin-Poole}}
\affiliation{Department of Physics and Astronomy, University of Waterloo, 200 University Ave W, Waterloo, ON N2L 3G1, Canada}

\author{G.~M.~Bernstein}
\affiliation{Department of Physics and Astronomy, University of Pennsylvania, Philadelphia, PA 19104, USA}

\author{\OrcidIDName{0000-0002-5636-233X}{A.~Choi}}
\affiliation{NASA Goddard Space Flight Center, 8800 Greenbelt Rd, Greenbelt, MD 20771, USA}

\author{M.~Gatti}
\affiliation{Department of Physics and Astronomy, University of Pennsylvania, Philadelphia, PA 19104, USA}

\author{\OrcidIDName{0000-0002-3730-1750}{G.~Giannini}}
\affiliation{Institut de F\'{\i}sica d'Altes Energies (IFAE), The Barcelona Institute of Science and Technology, Campus UAB, 08193 Bellaterra (Barcelona) Spain}
\affiliation{Kavli Institute for Cosmological Physics, University of Chicago, Chicago, IL 60637, USA}

\author{R.~A.~Gruendl}
\affiliation{Center for Astrophysical Surveys, National Center for Supercomputing Applications, 1205 West Clark St., Urbana, IL 61801, USA}
\affiliation{Department of Astronomy, University of Illinois at Urbana-Champaign, 1002 W. Green Street, Urbana, IL 61801, USA}

\author{M.~Jarvis}
\affiliation{Department of Physics and Astronomy, University of Pennsylvania, Philadelphia, PA 19104, USA}

\author{S.~Lee}
\affiliation{Jet Propulsion Laboratory, California Institute of Technology, 4800 Oak Grove Dr., Pasadena, CA 91109, USA}

\author{\OrcidIDName{0000-0001-9497-7266}{J. Mena-Fern{\'a}ndez}}
\affiliation{LPSC Grenoble - 53, Avenue des Martyrs 38026 Grenoble, France}

\author{A.~Porredon}
\affiliation{Centro de Investigaciones Energ\'eticas, Medioambientales y Tecnol\'ogicas (CIEMAT), Madrid, Spain}
\affiliation{Ruhr University Bochum, Faculty of Physics and Astronomy, Astronomical Institute, German Centre for Cosmological Lensing, 44780 Bochum, Germany}

\author{M.~Rodriguez-Monroy}
\affiliation{Instituto de Fisica Teorica UAM/CSIC, Universidad Autonoma de Madrid, 28049 Madrid, Spain}

\author{\OrcidIDName{0000-0002-1666-6275}{E.~Rozo}}
\affiliation{Department of Physics, University of Arizona, Tucson, AZ 85721, USA}

\author{\OrcidIDName{0000-0001-9376-3135}{E.~S.~Rykoff}}
\affiliation{SLAC National Accelerator Laboratory, Menlo Park, CA 94025, USA}
\affiliation{Kavli Institute for Particle Astrophysics \& Cosmology, P. O. Box 2450, Stanford University, Stanford, CA 94305, USA}

\author{\OrcidIDName{0000-0002-7187-9628}{T.~Schutt}}
\affiliation{Kavli Institute for Particle Astrophysics \& Cosmology, P. O. Box 2450, Stanford University, Stanford, CA 94305, USA}
\affiliation{SLAC National Accelerator Laboratory, Menlo Park, CA 94025, USA}
\affiliation{Department of Physics, Stanford University, 382 Via Pueblo Mall, Stanford, CA 94305, USA}

\author{E.~Sheldon}
\affiliation{Brookhaven National Laboratory, Bldg 510, Upton, NY 11973, USA}

\author{M.~A.~Troxel}
\affiliation{Department of Physics, Duke University Durham, NC 27708, USA}

\author{N.~Weaverdyck}
\affiliation{Department of Astronomy, University of California, Berkeley,  501 Campbell Hall, Berkeley, CA 94720, USA}
\affiliation{Lawrence Berkeley National Laboratory, 1 Cyclotron Road, Berkeley, CA 94720, USA}

\author{V.~Wetzell}
\affiliation{Department of Physics and Astronomy, University of Pennsylvania, Philadelphia, PA 19104, USA}

\author{M.~Aguena}
\affiliation{Laborat\'orio Interinstitucional de e-Astronomia - LIneA, Rua Gal. Jos\'e Cristino 77, Rio de Janeiro, RJ - 20921-400, Brazil}

\author{A.~Alarcon}
\affiliation{Institute of Space Sciences (ICE, CSIC),  Campus UAB, Carrer de Can Magrans, s/n,  08193 Barcelona, Spain}

\author{\OrcidIDName{0000-0002-7069-7857}{S.~Allam}}
\affiliation{Fermi National Accelerator Laboratory, P. O. Box 500, Batavia, IL 60510, USA}

\author{A.~Amon}
\affiliation{Department of Astrophysical Sciences, Princeton University, Peyton Hall, Princeton, NJ 08544, USA}

\author{F.~Andrade-Oliveira}
\affiliation{Department of Physics, University of Michigan, Ann Arbor, MI 48109, USA}

\author{\OrcidIDName{0000-0002-4687-4657}{J.~Blazek}}
\affiliation{Department of Physics, Northeastern University, Boston, MA, 02115, USA}

\author{\OrcidIDName{0000-0002-8458-5047}{D.~Brooks}}
\affiliation{Department of Physics \& Astronomy, University College London, Gower Street, London, WC1E 6BT, UK}

\author{\OrcidIDName{0000-0003-3044-5150}{A.~Carnero~Rosell}}
\affiliation{Instituto de Astrofisica de Canarias, E-38205 La Laguna, Tenerife, Spain}
\affiliation{Laborat\'orio Interinstitucional de e-Astronomia - LIneA, Rua Gal. Jos\'e Cristino 77, Rio de Janeiro, RJ - 20921-400, Brazil}
\affiliation{Universidad de La Laguna, Dpto. Astrofísica, E-38206 La Laguna, Tenerife, Spain}

\author{\OrcidIDName{0000-0002-3130-0204}{J.~Carretero}}
\affiliation{Institut de F\'{\i}sica d'Altes Energies (IFAE), The Barcelona Institute of Science and Technology, Campus UAB, 08193 Bellaterra (Barcelona) Spain}

\author{\OrcidIDName{0000-0002-7887-0896}{C.~Chang}}
\affiliation{Department of Astronomy and Astrophysics, University of Chicago, Chicago, IL 60637, USA}
\affiliation{Kavli Institute for Cosmological Physics, University of Chicago, Chicago, IL 60637, USA}

\author{\OrcidIDName{0000-0002-9745-6228}{M.~Crocce}}
\affiliation{Institut d'Estudis Espacials de Catalunya (IEEC), 08034 Barcelona, Spain}
\affiliation{Institute of Space Sciences (ICE, CSIC),  Campus UAB, Carrer de Can Magrans, s/n,  08193 Barcelona, Spain}

\author{L.~N.~da Costa}
\affiliation{Laborat\'orio Interinstitucional de e-Astronomia - LIneA, Rua Gal. Jos\'e Cristino 77, Rio de Janeiro, RJ - 20921-400, Brazil}

\author{M.~E.~S.~Pereira}
\affiliation{Hamburger Sternwarte, Universit\"{a}t Hamburg, Gojenbergsweg 112, 21029 Hamburg, Germany}

\author{\OrcidIDName{0000-0002-4213-8783}{T.~M.~Davis}}
\affiliation{School of Mathematics and Physics, University of Queensland,  Brisbane, QLD 4072, Australia}

\author{\OrcidIDName{0000-0002-0466-3288}{S.~Desai}}
\affiliation{Department of Physics, IIT Hyderabad, Kandi, Telangana 502285, India}

\author{\OrcidIDName{0000-0002-8357-7467}{H.~T.~Diehl}}
\affiliation{Fermi National Accelerator Laboratory, P. O. Box 500, Batavia, IL 60510, USA}

\author{\OrcidIDName{0000-0002-8446-3859}{S.~Dodelson}}
\affiliation{Department of Astronomy and Astrophysics, University of Chicago, Chicago, IL 60637, USA}
\affiliation{Kavli Institute for Cosmological Physics, University of Chicago, Chicago, IL 60637, USA}
\affiliation{Fermi National Accelerator Laboratory, P. O. Box 500, Batavia, IL 60510, USA}

\author{P.~Doel}
\affiliation{Department of Physics \& Astronomy, University College London, Gower Street, London, WC1E 6BT, UK}

\author{\OrcidIDName{0000-0001-8251-933X}{A.~Drlica-Wagner}}
\affiliation{Kavli Institute for Cosmological Physics, University of Chicago, Chicago, IL 60637, USA}
\affiliation{Fermi National Accelerator Laboratory, P. O. Box 500, Batavia, IL 60510, USA}
\affiliation{Department of Astronomy and Astrophysics, University of Chicago, Chicago, IL 60637, USA}

\author{A.~Fert\'e}
\affiliation{SLAC National Accelerator Laboratory, Menlo Park, CA 94025, USA}

\author{\OrcidIDName{0000-0003-4079-3263}{J.~Frieman}}
\affiliation{Kavli Institute for Cosmological Physics, University of Chicago, Chicago, IL 60637, USA}
\affiliation{Fermi National Accelerator Laboratory, P. O. Box 500, Batavia, IL 60510, USA}

\author{\OrcidIDName{0000-0002-9370-8360}{J.~Garc\'ia-Bellido}}
\affiliation{Instituto de Fisica Teorica UAM/CSIC, Universidad Autonoma de Madrid, 28049 Madrid, Spain}

\author{\OrcidIDName{0000-0001-9632-0815}{E.~Gaztanaga}}
\affiliation{Institute of Space Sciences (ICE, CSIC),  Campus UAB, Carrer de Can Magrans, s/n,  08193 Barcelona, Spain}
\affiliation{Institute of Cosmology and Gravitation, University of Portsmouth, Portsmouth, PO1 3FX, UK}
\affiliation{Institut d'Estudis Espacials de Catalunya (IEEC), 08034 Barcelona, Spain}

\author{\OrcidIDName{0000-0003-3270-7644}{D.~Gruen}}
\affiliation{University Observatory, Faculty of Physics, Ludwig-Maximilians-Universit\"at, Scheinerstr. 1, 81679 Munich, Germany}

\author{\OrcidIDName{0000-0003-0825-0517}{G.~Gutierrez}}
\affiliation{Fermi National Accelerator Laboratory, P. O. Box 500, Batavia, IL 60510, USA}

\author{W.~G.~Hartley}
\affiliation{Department of Astronomy, University of Geneva, ch. d'\'Ecogia 16, CH-1290 Versoix, Switzerland}

\author{\OrcidIDName{0000-0001-6718-2978}{K.~Herner}}
\affiliation{Fermi National Accelerator Laboratory, P. O. Box 500, Batavia, IL 60510, USA}

\author{S.~R.~Hinton}
\affiliation{School of Mathematics and Physics, University of Queensland,  Brisbane, QLD 4072, Australia}

\author{D.~L.~Hollowood}
\affiliation{Santa Cruz Institute for Particle Physics, Santa Cruz, CA 95064, USA}

\author{\OrcidIDName{0000-0002-6550-2023}{K.~Honscheid}}
\affiliation{Department of Physics, The Ohio State University, Columbus, OH 43210, USA}
\affiliation{Center for Cosmology and Astro-Particle Physics, The Ohio State University, Columbus, OH 43210, USA}

\author{\OrcidIDName{0000-0001-6558-0112}{D.~Huterer}}
\affiliation{Department of Physics, University of Michigan, Ann Arbor, MI 48109, USA}

\author{\OrcidIDName{0000-0001-5160-4486}{D.~J.~James}}
\affiliation{Center for Astrophysics $\vert$ Harvard \& Smithsonian, 60 Garden Street, Cambridge, MA 02138, USA}

\author{\OrcidIDName{0000-0001-8356-2014}{E.~Krause}}
\affiliation{Department of Astronomy/Steward Observatory, University of Arizona, 933 North Cherry Avenue, Tucson, AZ 85721-0065, USA}

\author{\OrcidIDName{0000-0003-0120-0808}{K.~Kuehn}}
\affiliation{Australian Astronomical Optics, Macquarie University, North Ryde, NSW 2113, Australia}
\affiliation{Lowell Observatory, 1400 Mars Hill Rd, Flagstaff, AZ 86001, USA}

\author{\OrcidIDName{0000-0002-1134-9035}{O.~Lahav}}
\affiliation{Department of Physics \& Astronomy, University College London, Gower Street, London, WC1E 6BT, UK}

\author{\OrcidIDName{0000-0003-0710-9474}{J.~L.~Marshall}}
\affiliation{George P. and Cynthia Woods Mitchell Institute for Fundamental Physics and Astronomy, and Department of Physics and Astronomy, Texas A\&M University, College Station, TX 77843,  USA}

\author{\OrcidIDName{0000-0002-6610-4836}{R.~Miquel}}
\affiliation{Instituci\'o Catalana de Recerca i Estudis Avan\c{c}ats, E-08010 Barcelona, Spain}
\affiliation{Institut de F\'{\i}sica d'Altes Energies (IFAE), The Barcelona Institute of Science and Technology, Campus UAB, 08193 Bellaterra (Barcelona) Spain}

\author{\OrcidIDName{0000-0002-7579-770X}{J.~Muir}}
\affiliation{Perimeter Institute for Theoretical Physics, 31 Caroline St. North, Waterloo, ON N2L 2Y5, Canada}

\author{J.~Myles}
\affiliation{Department of Astrophysical Sciences, Princeton University, Peyton Hall, Princeton, NJ 08544, USA}

\author{\OrcidIDName{0000-0001-9186-6042}{A.~Pieres}}
\affiliation{Laborat\'orio Interinstitucional de e-Astronomia - LIneA, Rua Gal. Jos\'e Cristino 77, Rio de Janeiro, RJ - 20921-400, Brazil}
\affiliation{Observat\'orio Nacional, Rua Gal. Jos\'e Cristino 77, Rio de Janeiro, RJ - 20921-400, Brazil}

\author{\OrcidIDName{0000-0002-2598-0514}{A.~A.~Plazas~Malag\'on}}
\affiliation{SLAC National Accelerator Laboratory, Menlo Park, CA 94025, USA}
\affiliation{Kavli Institute for Particle Astrophysics \& Cosmology, P. O. Box 2450, Stanford University, Stanford, CA 94305, USA}

\author{J.~Prat}
\affiliation{Nordita, KTH Royal Institute of Technology and Stockholm University, Hannes Alfv\'ens v\"ag 12, SE-10691 Stockholm, Sweden}
\affiliation{Department of Astronomy and Astrophysics, University of Chicago, Chicago, IL 60637, USA}

\author{M.~Raveri}
\affiliation{Department of Physics, University of Genova and INFN, Via Dodecaneso 33, 16146, Genova, Italy}

\author{S.~Samuroff}
\affiliation{Department of Physics, Northeastern University, Boston, MA 02115, USA}
\affiliation{Institut de F\'{\i}sica d'Altes Energies (IFAE), The Barcelona Institute of Science and Technology, Campus UAB, 08193 Bellaterra (Barcelona) Spain}

\author{\OrcidIDName{0000-0002-9646-8198}{E.~Sanchez}}
\affiliation{Centro de Investigaciones Energ\'eticas, Medioambientales y Tecnol\'ogicas (CIEMAT), Madrid, Spain}

\author{\OrcidIDName{0000-0003-3054-7907}{D.~Sanchez Cid}}
\affiliation{Centro de Investigaciones Energ\'eticas, Medioambientales y Tecnol\'ogicas (CIEMAT), Madrid, Spain}

\author{\OrcidIDName{0000-0002-1831-1953}{I.~Sevilla-Noarbe}}
\affiliation{Centro de Investigaciones Energ\'eticas, Medioambientales y Tecnol\'ogicas (CIEMAT), Madrid, Spain}

\author{\OrcidIDName{0000-0002-3321-1432}{M.~Smith}}
\affiliation{Physics Department, Lancaster University, Lancaster, LA1 4YB, UK}

\author{\OrcidIDName{0000-0002-7047-9358}{E.~Suchyta}}
\affiliation{Computer Science and Mathematics Division, Oak Ridge National Laboratory, Oak Ridge, TN 37831}

\author{\OrcidIDName{0000-0003-1704-0781}{G.~Tarle}}
\affiliation{Department of Physics, University of Michigan, Ann Arbor, MI 48109, USA}

\author{\OrcidIDName{0000-0001-7211-5729}{D.~L.~Tucker}}
\affiliation{Fermi National Accelerator Laboratory, P. O. Box 500, Batavia, IL 60510, USA}

\author{\OrcidIDName{0000-0002-7123-8943}{A.~R.~Walker}}
\affiliation{Cerro Tololo Inter-American Observatory, NSF's National Optical-Infrared Astronomy Research Laboratory, Casilla 603, La Serena, Chile}

\author{P.~Wiseman}
\affiliation{School of Physics and Astronomy, University of Southampton,  Southampton, SO17 1BJ, UK}

\author{Y.~Zhang}
\affiliation{Cerro Tololo Inter-American Observatory, NSF's National Optical-Infrared Astronomy Research Laboratory, Casilla 603, La Serena, Chile}



\begin{abstract}
Synthetic source injection (SSI), the insertion of sources into pixel-level on-sky images, is a powerful method for characterizing object detection and measurement in wide-field, astronomical imaging surveys. Within the Dark Energy Survey (DES), SSI plays a critical role in characterizing all necessary algorithms used in converting images to catalogs, and in deriving quantities needed for the cosmology analysis, such as object detection rates, galaxy redshift estimation, galaxy magnification, star-galaxy classification, and photometric performance. We present here a source injection catalog of $146$ million injections spanning the entire $5000 \,\,{\rm deg}^2$ DES footprint, generated using the \Balrog SSI pipeline. Through this SSI sample, we demonstrate that the DES Year 6 (Y6) image processing pipeline provides accurate estimates of the object properties, for both galaxies and stars, at the percent-level, and we highlight specific regimes where the accuracy is reduced. We then show the consistency between SSI and data catalogs, for all galaxy samples developed within the weak lensing and galaxy clustering analyses of DES Y6. The consistency between the two catalogs also extends to their correlations with survey observing properties (seeing, airmass, depth, extinction, etc.). Finally, we highlight a number of applications of this catalog to the DES Y6 cosmology analysis, such as estimates of the redshift distribution and lens magnification. This dataset is the largest SSI catalog produced at this fidelity and will serve as a key testing ground for exploring the utility of SSI catalogs in upcoming surveys such as the Vera C. Rubin Observatory Legacy Survey of Space and Time.
\end{abstract}



\section{Introduction}\label{sec:intro}

Photometric datasets have a storied history spanning more than a century:\footnote{Photometry performed \textit{through photography/plates} is a little over a century old, whereas photometry-by-eye has existed for over two millenia and extends back to the catalogs of Hipparchus of Nicea in 120 BC and Ptolemy's ``Algamest'' from 137 AD \citep{Miles:2007:History}.} from the popularization of long-exposure photographic plates by \citet{Roberts:1893}, to the half-sky survey of \citet{Barnard:1919, Barnard:1927} revealing the faint structures (``dark nebulae'') of the Milky Way, to the international Carte du Ciel project (circa 1887) that spanned telescopes from twenty institutions and whose measurements of stars continue to be used today for long-baseline astrometry \citep{Lehtinen:2023:CarteduCiel}. The first hints of the large-scale structure of our Universe were also discovered through photometric surveys: from the work of \citet{Hubble:1934} in constructing photometric galaxy (``nebulae'') catalogs from $650 \deg^2$ and uncovering signatures of their spatial clustering, to that of \citet{Abell:1958} in discovering and cataloging large clusters of galaxies.\footnote{While this entire description mentions (a subset of) the history of \textit{photographic} surveys --- i.e. those that used photographic plates to image objects --- we point out the first precursors to our modern widefield surveys were the visually identified, non-photographic catalogs of Sir William Herschel \citep{Herschel:1786, Herschel:1789, Herschel:1802}, which covered thousands of square degrees of the sky and were eventually integrated into the NGC catalogs \citep{Dreyer:1888:NGC} that are still referred to today.} Wide-field, digital photometric surveys, a more recent endeavor in this long history, have amplified the relevance of such datasets \citep[\eg][]{York_2000, Flaugher_2005, deJong_2013, Aihara_2018, Dawson_2013, Dawson_2016}.

A pertinent point is that analyses of all above/forthcoming datasets rely on knowledge of the quantity $P(\Sigma_{\rm meas} \,|\, \Sigma_{\rm true}, X\ldots)$, i.e. the probability distribution of observed properties ($\Sigma_{\rm meas}$) of a certain astronomical object given the object's true properties ($\Sigma_{\rm true}$), and given characterizations of the survey/image properties ($X$) such as seeing, sky background etc. Modern pipelines for photometric measurements include a wide variety of steps \citep[\eg][]{Stoughton_2002, Magnier:2006:PanstarrsProcessing, Skrutskie:2006:2MASSProccessing, Valdes_2014, DeJong:2015:KidsProcessing, Bosch:2018:Hscprocessing, DESDM_ImageProcessingPipeline, Wright:2024:Kidsprocessing} that are chosen/designed to make the distribution $P(\Sigma_{\rm meas} \,|\, \Sigma_{\rm true}, X\ldots)$ as narrow and unbiased\footnote{In this context, an ``unbiased'' distribution is one that minimizes the difference $\Sigma_{\rm meas} - \Sigma_{\rm true}$.} as possible. Reliable cosmological inference now requires a robust calibration of this distribution, by accounting for the complexities in the imaging data (such as, for example, the impact of a range of purely observational effects) and for the algorithms in the image processing pipelines.\footnote{For example, the relative photometric accuracy in DES Y6 is calibrated to better than 1\% \citep{Y6Gold}, and the morphological measurements (primarily the galaxy ellipticity/orientations) are also better than 1\% \citep{Y6Mdet, Y6Imsims}.} 

Synthetic source injection (SSI) is a principled, data-driven approach for calibrating this distribution, $P(\Sigma_{\rm meas} \,|\, \Sigma_{\rm true}, X\ldots)$, which we will henceforth refer to as the ``transfer function''. SSI begins from a catalog of astronomical objects with measurements that have significantly lower noise relative to the actual survey that SSI is being run on; these can either be purely simulated sources with noiseless measurements, or sources from regions observed to better depth than the main survey and therefore, with measurements that have significantly lower noise than those from the main survey. By injecting these sources with known properties (such as size, luminosity, color, etc.) into the on-sky images from the main survey, and processing those augmented images through the same pipeline as the survey data, we can understand the performance of the image processing pipeline in characterizing different kinds of astronomical sources. Such a calibration is critical for many downstream analyses of photometric datasets (we describe the analyses relevant for our work later in this paper). 

By injecting such sources into the real pixel-level CCD images, the SSI catalog traces the complexities of the on-sky data in a way that is challenging to replicate, at full fidelity, with purely simulated images. Complexities include instrument signatures, spatially varying backgrounds, blending with complicated scenes of real galaxies etc. The method naturally also incorporates the exact distributions of delivered image quality, photometric zeropoints, sky background variances, etc., corresponding to the images used in deriving the DES Y6 data products and the corresponding cosmology results.

For these reasons, SSI has been used in many analyses within the Dark Energy Survey \citep[DES,][]{Flaugher_2005}: see  \citet[][henceforth, \citetalias{Balrog_Y1}]{Balrog_Y1} and \citet[][henceforth, \citetalias{BalrogY3}]{BalrogY3} for the DES Year 1 (Y1) and Year 3 (Y3) SSI pipelines, respectively. The DES Y1 SSI campaign emulated some pieces of the full image processing pipeline, while the DES Y3 SSI campaign emulated nearly all pieces of this pipeline. In this work, we describe the DES Y6 SSI campaign, which further enhances the consistency between the image processing used in the SSI pipeline (denoted as \Balrog) and that used on the data. Our injection catalog is taken from \citep[][henceforth, \citetalias{DES_DF_Y3}]{DES_DF_Y3} and comprises fits to the light profiles of objects in the DES Y3 deep fields --- where these measurements have significantly lower noise relative to those from the main survey  --- and is therefore completely data-driven. The main survey covers $5000 \deg^2$ of the sky in the $griz$ photometric bands. See \citet{Y6Gold} for a detailed characterization of this dataset.

Other surveys are also undertaking SSI to understand the statistical properties and biases of their datasets: for example, the \textsc{Obi-wan} pipeline \citep{Kong:2024:ObiWan} used in the Dark Energy Spectroscopic Instrument \citep{DESI:2016:Part1}, the \textsc{SynPipe} software used in the Hyper Suprime-Cam (HSC) data \citep{Miyazaki:2018:HSC}, the \textsc{CosmoDC} datasets \citep{Sanchez:2020:CosmoDC2} used in the Dark Energy Science Collaboration \citep[DESC,][]{DESC:2018:SRD}, or the \Balrog-equilavent in the DECam All Data Everywhere (DECADE) cosmic shear project \citep{paper1, paper3}.

This work supports the DES Y6 weak lensing and galaxy clustering cosmology analysis by presenting the associated SSI dataset, its validations, and its applications. We detail the source injection pipeline, the image processing pipeline, and all updates made to both in relation to the previous Y3 effort. Of particular note is a new, optimized injection scheme that improves the number of usable injections by up to a factor of 12, and also an extension of SSI to the full, \(\sim\)$5000\deg^2$ footprint of the survey. As a result, the Y6 SSI catalog has nearly two orders of magnitude more sources compared to the Y3 SSI catalog. We also characterize the performance of the image processing pipeline in accurately recovering the input properties of simulated objects, and then showcase the similarities between the synthetic samples and the actual catalogs from the Y6 datasets. The Y3 analysis \citepalias{BalrogY3} also presented a detailed study of their photometric measurements, identifying failure modes (\eg blending, sky background errors etc.). Given those algorithms studied in \citetalias{BalrogY3} are being used in DES Y6 with minor/no changes, we do not repeat all aspects of their study in this work.

The \Balrog synthetic dataset plays critical roles in estimating multiple, key quantities for the cosmology analysis: for example, estimating the redshift distributions of the different galaxy samples \citep{Myles:2021:SOMPZ, Sanchez:2023:highz, Giannini:2024:DES, paper2}, the magnification of galaxies due to foreground structure \citep{Garcia-Fdez:2018:Magnification, Magnification_Y3}, and validating the systematics treatment of the galaxy catalogs \citep{Y3_GalClustering}. We describe a few of these applications in this work, with a focus on how \Balrog is employed in these analyses.

This work is organized as follows: in Sections \ref{sec:balrog} and \ref{sec:y6} we detail the image processing pipeline of Y6 \Balrog, including all similarities/differences with the fiducial pipeline applied to the data and with the previous Y3 \Balrog pipeline. In Section \ref{sec:photo} we validate the synthetic samples, through a number of comparisons with the true, injected samples, and also with the actual data catalogs, and then examine the photometric accuracy of the synthetic sample. In Section \ref{sec:applications} we detail some key applications of \Balrog in the Y6 cosmology efforts, as detailed above. We conclude in Section \ref{sec:conclusion}. In Appendix \ref{appx:SPmaps_Lens} we describe the correlations between the galaxy number density of two lens samples and the survey property  maps. Appendix \ref{appx:performancestar} details further photometric validations for the synthetic star sample in \Balrog.

\section{The Balrog Pipeline}\label{sec:balrog}

\Balrog --- the DES internal nomenclature for our SSI pipeline\footnote{To paraphrase the footnote from the original introduction of \Balrog in \citet{Balrog_Y1}: ``\textit{\Balrog is \textit{not} an acronym. The software was born out of the authors digging too greedily and too deep into their data, ergo the name.}''} --- injects sources with known photometry and morphology into on-sky images, and then replicates the entire processing of the DES Data Management pipeline \citep[DESDM;][]{DESDM_ImageProcessingPipeline}. The processed images are then run through the same postprocessing steps as the data to obtain a variety of object catalogs. These catalogs are then position-matched to the true, injected sources to thereby obtain a distribution of noisy, realistic measurements for each injected source.

We now detail the full pipeline, including both image processing and galaxy catalog generation, in Section \ref{sec:DESDM_Pipeline}. We describe specific differences between the \Balrog and DESDM pipelines in Section \ref{sec:DESDM_differences}, with explicit justifications for each difference.

\subsection{Emulating the DES Data Management (DESDM) Pipeline}\label{sec:DESDM_Pipeline}

The image processing in DES follows the methods presented in \citet{DESDM_ImageProcessingPipeline}, with some updates made during the Y3 \citep{Y3Gold} and Y6 analyses \citep{Y6Gold}. The \Balrog pipeline replicates nearly all steps perfomed by DESDM. Some minor steps are avoided in order to reduce the complexity of the pipeline, to speed up the execution, or due to other limitations. All relevant differences are detailed in Section \ref{sec:DESDM_differences}.

The primary unit of the DESDM coadd image processing is the ``coadd tile'' \citep{DESDM_ImageProcessingPipeline}, which is a $0.7 \times 0.7 \deg^2$ square region of the sky, spanning $10,000 \times 10,000$ pixels. The DES footprint is partitioned into 10,169 tiles, all of which are re-processed with the \Balrog injections; see Figure \ref{fig:footprint} for the area covered by the \Balrog sample. DES accumulated a total of 83,706 exposures in the wide survey, with each location in the footprint being imaged an average of 8 times in each of the $grizY$ bands \citep{DES_DR2}. The $Y$ band is not used in the cosmology analysis, while the $g$ band is not used in object detection. The image processing pipeline performs coaddition on all exposures that overlap with the defined tile region, and generates catalogs for the tile.

The \Balrog pipeline for an individual coadd tile undertakes the following steps:

\begin{itemize}
    \item[1.] \textbf{Database query:} Given a coadd tile, we find all CCD images, in the $griz$ bands, that are associated with that tile; that is, all images used in generating the coadd image of that tile. These data products are hosted at the National Center for Supercomputing Application (NCSA) and queried locally for processing. Note that this step involves locating not only the raw images, but also models of the image point spread function (PSF) from \textsc{Piff} \citep{PIFF_Y3}\footnote{\textsc{Piff} is a more recent addition to the DES processing pipeline: it was not in the original pipeline of \citet{DESDM_ImageProcessingPipeline} but was later developed and integrated into the pipeline to improve the PSF modelling (which in turn improves the galaxy shape estimation needed for weak-lensing cosmology).}, astrometric solutions from \textsc{Scamp} \citep{SCAMP}, sky background, noise, variance, and mask plane images from \SE \citep{SourceExtractor}, etc.
    
    \item[2.] \textbf{Null-weight images:} The actual images used for generating coadd images (as well as other imaging products therein) are not the raw images downloaded in Step 1 above, but rather \textit{null-weighted} images. These images have had an initial detrending and instrument signature removal applied via the DESDM Final Cut pipeline \citep{DESDM_ImageProcessingPipeline}. This includes a masking and interpolation step to remove problematic features (e.g., cosmic rays) in the images, and also corrects for biases and non-linear responses in the CCD pixels \citep[][see their Section 3.3]{DESDM_ImageProcessingPipeline}. The null-weighted CCD images are used as the starting point for the source injection step in \Balrog.

    \item[3.] \textbf{Source injection into CCD images:} Synthetic sources are injected into all CCD images from Step 2. The object light profiles account for the pixel scale and include shot noise. The catalog of injections we use is obtained from the DES deep-field data, and is described in Section \ref{sec:SourceCat}. The injection models are ingested and rendered using a combination of the \textsc{Ngmix} \citep{Sheldon:2015:Ngmix} and \textsc{Galsim} packages \citep{galsim}. We convolve the galaxy models using the \textsc{Piff} PSF \citep{PIFF_Y3}, evaluated at the location of the injected source. See \citet{Y6PSF} for details on the Y6 PSF model. All galaxies are rotated randomly prior to injection. The source photometry is reddened as $m_{\rm red} = m_{\rm true} + E(B-V)R_b$ where the $E(B-V)$ value is the average extinction over a given CCD image, calculated using the map of \citep{Schlegel_1998}, and $R_b$ is a band-dependent coefficient for each of the $griz$ filters, provided in \citet[][see their Section 4.2.3]{DES_DR2}.\footnote{We show below, in Section \ref{sec:sec:sec:sysmaps}, that \Balrog matches the data in capturing the spatial correlations between the galaxy catalogs and the dust extinction map.} This follows the same procedure done in \Balrog from DES Y3. After this, the object's light profile is added to the CCD image at the relevant location (see Section \ref{sec:InjWgtscheme} for more details). In Y6, we also utilize multiple modes of source injection that aid different science goals within the Y6 cosmology analysis: see Section \ref{sec:InjWgtscheme} and Section \ref{sec:FidAndMag}.

    \item[4.] \textbf{Generate coadd image with \textsc{Swarp}}: Step 3 provides us a set of new ``null-weighted'' images that include real sources \textit{and} the synthetic sources. These are passed into the \textsc{Swarp} \citep{SWARP} routine to produce the coadd image. A coadd is generated for each band, and the coadds from the $riz$ bands are further coadded together to obtain a ``detection'' coadd. This latter image is used for source detection. While the $g$ band is simulated in \Balrog and used in the measurements, it is not used for detection following the  procedure done to the data.

    \item[5.] \textbf{Source catalogs and \textsc{MEDS} files}: The detection coadd image from Step 4 is analyzed with \SE \citep{SourceExtractor} to obtain the detection catalog for the given coadd tile. We run \SE in ``dual mode'' to generate photometric measurements in each of the $griz$ bands for all detections. We then make Multi-Epoch Data Structure \citep[\textsc{MEDS};][]{MEDS} files, which consistently collate all images of an object (in small, square cutouts of at least $32 \times 32$ pixels\footnote{The pixel scale of a coadd image is, to good approximation, $0.263$ arcseconds.} or more) from all exposures/CCDs and all bands, for all detected objects. These \textsc{MEDS} files are used in generating almost all of the measurements for galaxy samples that we discuss; the notable exception is the \mdet sample (defined in point 8 below), which requires a slightly different detection and \textsc{MEDS}-making procedure.

    \item[6.] \textbf{\fitvd catalogs:} The first catalog we generate --- that serves as a general purpose photometric measurement in DES, and also as a base sample from which we derive lens samples used in the galaxy clustering analysis --- are the \fitvd fits, named after the codebase used to derive them.\footnote{\url{https://github.com/esheldon/fitvd}} During fitting of a given object $A$, all pixels in its \textsc{MEDS} data are checked and we mask any pixel whose closest detection is not object $A$. This is done using the segmentation map provided by \SE and collected in the \textsc{MEDS} data. In DES Y3, we also had a multi-object fitting routine (MOF) that jointly fit groups of galaxies (rather than one galaxy at a time) to handle blending effects. This measurement was foregone in Y6 given it is computationally expensive, and was ultimately not used for cosmology analyses in Y3.
   
    \item[7.] \textbf{Cell-based coadds:} In DES Y6, we utilize the \mdet algorithm \citep{Sheldon:2020:Mdet, Sheldon:2023:Mdet} for estimating the shear of galaxies. This algorithm does not work on the coadds built in Step 4 and instead generates its own coadds, denoted as ``cell-based'' coadds. The latter are smaller than the coadd tiles we discuss here, and are constructed to have a continuous, well-defined PSF model. These cell-based coadds are generated using the \textsc{Pizzacutter} codebase.\footnote{\url{https://github.com/beckermr/pizza-cutter}} See \citet{Y6Mdet} for more details on \mdet and the cell-based coadds.

    \item[8.] \textbf{\mdet source sample:} The \mdet estimator generates five versions of the cell-based coadds in step 7. The first is the original data, and the remaining four have an artificial shear applied to the full cell-based coadd in different directions, to empirically calibrate the response of object properties to external shear \citep[\eg][]{Sheldon:2020:Mdet, Sheldon:2023:Mdet, Y6Mdet}. The \SE detection algorithm is run on all five images,\footnote{For the \mdet step alone, we match the choices of \citet{Y6Mdet} and use the \textsc{sep} \citep{sep} package, which is a python-based wrapper of \SE.} thereby providing five detection catalogs. We estimate shapes for all objects in each of the five catalogs, once again using the \textsc{Ngmix} package. This serves as one of our ``source'' samples, which is used for measuring weak lensing.

    \item[9.] \textbf{Bayesian Fourier Domain (BFD) source sample:} BFD \citep{Bernstein:2014:BFD} is a second estimator of the weak-lensing shear of a galaxy. For this method, we only need the \textsc{MEDS} files built in Step 5. The BFD pipeline includes two steps: measuring the moments of the galaxy, and performing a bayesian integration of the moments to assign a shear estimate to the galaxy. We only perform the first step within the \Balrog pipeline. See \citet{Y6BFD} for more details.

    \item[10.] \textbf{Catalog matching:} Finally, for each type of catalog we generate --- \fitvd \footnote{All other catalogs used in this work --- BFD, \maglim, \redmagic --- use a subset of the detections present in the \fitvd catalog and therefore have the same matching as the latter catalog.} (step 6) and \mdet (step 8) --- we match the resulting output catalogs to the injection truth catalogs. For each synthetic source, injected at a given sky position $x = \{{\alpha_{2000}, \delta_{2000}}\}$, we find the nearest object in the output catalog. If the distance to the nearest object is within $<0.5\arcsec$ it is considered a successful match. More details on the matching process are also given in Section \ref{sec:Y3toY6}. This matching is done separately for the \fitvd and \mdet catalogs. For \mdet this procedure is repeated for all five detection catalogs.

    \item[11.] \textbf{\textsc{Maglim++} lens sample:} The \textsc{Maglim++} catalog \citep{Y6Lens} is one of the lens samples\footnote{Following existing DES works, we use the term ``lens sample'' to denote a sample used for making galaxy clustering measurements.} in the cosmology analysis and its selection requires a preliminary redshift estimate of the galaxies. Thus, we take all synthetic objects (after the postprocessing mentioned above) and we assign a redshift to each using the Directional Neighborhood Fitting (DNF) algorithm \citep{DNF}. The inputs to this are the magnitudes from the \fitvd catalog. The \textsc{Maglim++} sample is selected using the DNF redshifts as well as color selections performed using photometry from both \fitvd and from   measurements in \textsc{WISE} infrared survey \citep{Wright:2010:WISE}. \Balrog does not simulate the \textsc{WISE} photometry, and is instead limited to the DES photometric bands. However, the deep-field galaxies we inject have associated \textsc{WISE} photometry, so all wide-field realizations of a given deep-field galaxy are assigned the \textsc{WISE} photometry estimated for that deep-field galaxy. While this is an approximation, we find that the synthetic \maglim catalog still accurately reproduces the data. For example, it exhibits spatial correlations with survey properties (including the \textsc{WISE} magnitude limit) that are consistent with the data catalog (see Figure \ref{fig:survey_properties_maglim}).

    \item[12.] \textbf{\textsc{RedMaGic} lens sample:} The \textsc{RedMaGic} sample \citep{Redmagic} is a second lens sample for the cosmology analysis, but is also used in constructing the redshift distributions of other galaxy samples \citep[\eg][]{GattiGiannini:2022:Wz, Giannini:2024:DES, Y6WZ}. In DES Y6, we use it for informing the redshift distributions of the \mdet, BFD, and \maglim samples. The sample is constructed from color-based selections, in order to identify red galaxies whose redshifts can be estimated with higher accuracy. The inputs to the algorithm are the \fitvd magnitudes (and magnitude errors) in addition to survey observing conditions, such as the sky background estimates mentioned in Section \ref{fig:survey_properties}.
\end{itemize}

All pipelines for image processing and catalog generation are explicitly version-matched to follow those used in the processing of the actual data. Therefore \Balrog uses the exact software as the data.

\subsection{Differences from DESDM}\label{sec:DESDM_differences}

Not all steps from the DESDM pipeline are reproduced in \Balrog. We highlight here four relevant steps that are simplified in \Balrog, and discuss the motivation for altering them.

\begin{itemize}
    \item[1.] \textbf{PSF estimation:} For all data processing of a given image, we use the \textsc{Piff} model associated with a given CCD image (see \citet{Y6PSF} for details on the Y6 PSF model), and do not remeasure them after injecting our sources. This makes the approximation that the PSF model of the image is unchanged with/without the synthetic sources. Our choice follows the same approach used in all \Balrog and image simulation pipelines within DES Y1 and Y3 \citep[\citetalias{Balrog_Y1, BalrogY3};][]{Y3Imsims}, and we have no indication (from \Balrog catalogs of previous years or from the catalog in Y6) that this approximation causes artifacts in the catalog. Thus, the same, fiducial PSF model is used for the SSI step and for all detection/measurement steps that follow.

    \item[2.] \textbf{Astrometric solutions:} In the DESDM processing pipeline, all single-epoch exposures are fit with an astrometric solution. However, during the coadd processing, all CCD images belonging to a coadd tile (which necessarily includes images from different exposures) are jointly fit using \textsc{Scamp} \citep{SCAMP} to determine a new astrometric solution. In \Balrog we take the existing \textsc{Scamp} solutions of all images contributing to the coadd tile and do not run \textsc{Scamp} again during our coaddition of the source-injected images. Similar to point (1), this approximation is used in all \Balrog and image simulation pipelines, with no indication of deficiencies in the data quality due to skipping this step.

    \item[3.] \textbf{Photometric calibration}: the photometric calibration (i.e., zeropoint of each CCD image) is not recomputed in the SSI pipeline. The value of the zeropoint is primarily sensitive to instrument response and atmospheric conditions \citep{FGCM}, and not on the number density of sources. We use the zeropoints estimated for the original CCD images to compute the signal amplitude of injected sources, and therefore, to avoid circularity, we do not repeat the zeropoint calculation on the source-injected images.

    \item[4.] \textbf{Reduced \SE parameter list}: Finally, we do not generate the exact same list of \SE properties for objects as is done in the data. For example, properties associated with the PSF are skipped. This choice is consistent with what is done in \Balrog for DES Y3, and is motivated by selecting only those properties that are needed for the key DES cosmology science goals. The reduction in number of output properties greatly reduces the computational expense of the \SE step, and allows us to generate synthetic samples for a larger set of tiles (i.e., the full DES footprint), which in turn improves all DES science cases.
\end{itemize}

In general, \Balrog does not redo any of the ``single-epoch'' processing and only performs the ``multi-epoch'' processing; see Figure 3 (and text therein) from \citet{DESDM_ImageProcessingPipeline} for more details on these two classes of pipelines.

\section{Balrog in DES Year 6}\label{sec:y6}

\begin{figure}
    \centering
    \includegraphics[width = \columnwidth]{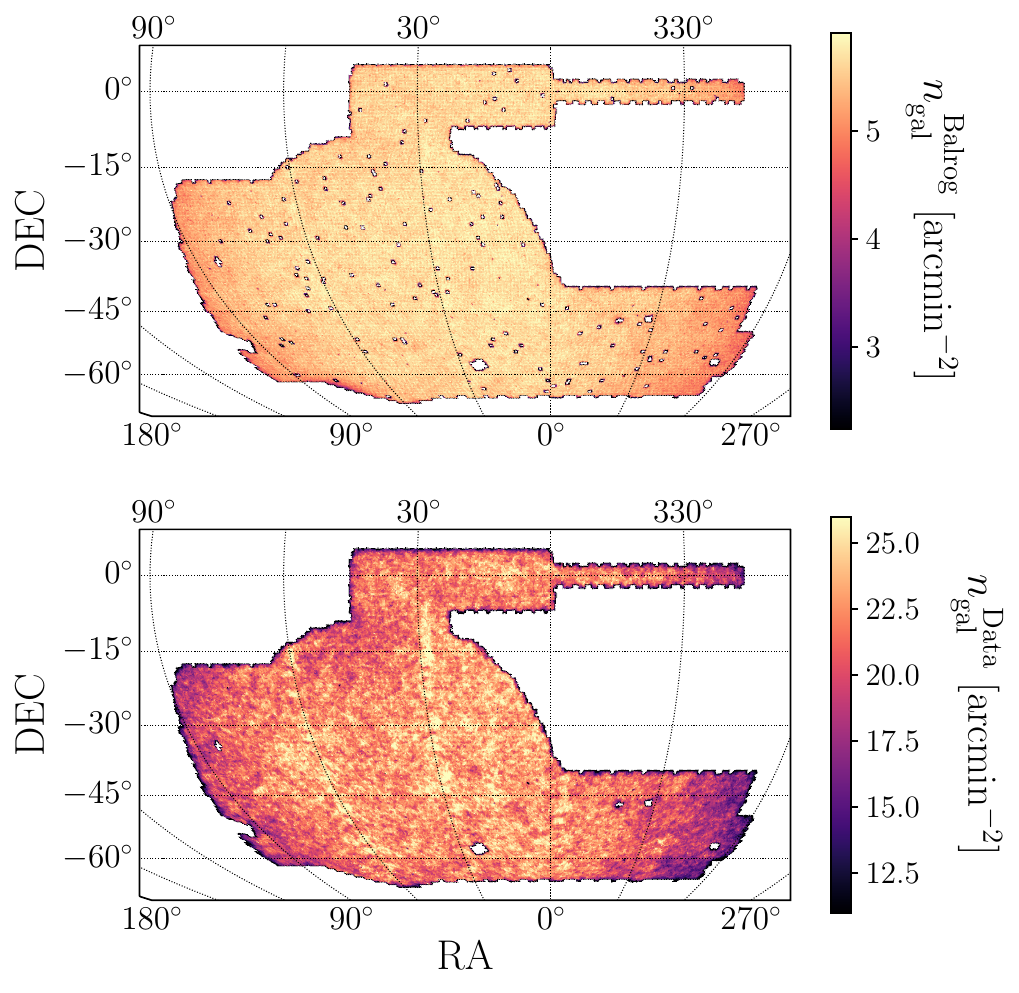}
    \caption{\textbf{Top:} the number density of all detections in the Y6 balrog sample, using only high purity galaxies of the Gold catalog (defined with the star-galaxy classifier \texttt{EXT\_MASH = 4}; see \citet[][Section 4.2]{Y6Gold} for more details). The isolated, lower-density locations correspond to a (random) handful of failed tiles in the \Balrog run, contributing $<1\%$ of area. \textbf{Bottom:} the same but for the actual DES Y6 data. The \Balrog sample covers the entire DES imaging area. The density in Balrog is lower than data by construction, and is done to prevent blending of injected sources.}
    \label{fig:footprint}
\end{figure}

A number of updates have occurred in the \Balrog pipeline between the DES Y3 analysis and the current Y6 analysis. These include changes to the source injection catalog (Section \ref{sec:SourceCat}), the injection schemes (Section \ref{sec:InjWgtscheme}), and the specific runs (and variants) performed in \Balrog (Section  \ref{sec:FidAndMag}). We also highlight other smaller, but salient, differences in Section \ref{sec:Y3toY6}.

The runs were performed primarily at the \textsc{FermiGrid} computing center at Fermilab, and also at the National Energy Research Science Computing (NERSC) system. The total computing cost of all \Balrog runs was approximately eight million CPU hours, split between two million CPU hours for 7000 tiles run at Fermilab, and six million CPU hours for 5000 tiles run at NERSC. The differences in the required CPU hours per tile reflect the differences in the configurations of the two distributed computing systems. Some parts of the pipeline are not designed to use all available processors (for example, the processing pipeline uses a single-threaded version of \SE) and thus have CPU loads that can change depending on the system.

\subsection{Injection catalog}\label{sec:SourceCat}

In Y3 \Balrog, the injection catalog was obtained from \citetalias{DES_DF_Y3} and provides \fitvd models of galaxies detected in the DES Y3 deep fields. This is a two-component model --- using Sersic profiles of index $n = 4$ and $n = 1$ for the bulge and disk components, respectively. The ratio of the total flux contained in the two components is given a Gaussian prior with mean, $\mu = 0.5$, and width $\sigma = 0.1$. The two components share the same half-light radius and ellipticity parameters in the fit.

A similar, but slightly different, approach is done in Y6 \Balrog: we continue using the same deep-field detections, but the galaxy models were re-fit using the latest version of the \fitvd algorithm, to be consistent with the Y6 processing. We use the de-reddenned fluxes of these objects during injection, where the de-reddenning is performed using the same image-by-image correction as done in Y3 Balrog \citepalias[][see their section 3.1.1]{BalrogY3}. We show in Figure \ref{fig:survey_properties} that our synthetic galaxies show correlations with the extinction map that are consistent with that found in the data catalogs.

Our injection catalog is obtained by taking the catalog of \citetalias{DES_DF_Y3} and then applying the following selections, which are the same as those motivated in \citetalias{BalrogY3} (see their Section 3.4):

\begin{align}\label{eqn:DFSampleCuts}
     & \texttt{in\_VHS\_footprint} == 1 \nonumber\\
     \texttt{AND} \,\,\,\,& \texttt{flags} == 0 \nonumber\\
     \texttt{AND} \,\,\,\,& \texttt{mask\_flags} == 0 \nonumber\\
     \texttt{AND} \,\,\,\,& \texttt{SNR\_{griz}} > -3 \nonumber\\
     \texttt{AND} \,\,\,\,& \texttt{avg\_bdf\_mag} < 25.4 \nonumber\\
     \texttt{AND} \,\,\,\,& \texttt{bdf\_T} < 100\,\,\texttt{arcsec}^2
\end{align}

All selections above are made using quantities found in (or derived from) the DES Y3 deep-field catalog. The first three selections remove all objects with problematic fits or objects in regions of the deep fields that do not have the multi-band coverage required by some applications of \Balrog (see Section \ref{sec:sec:SOMPZ}). The \texttt{SNR} selection removes objects that were determined to be fit failures during an inspection done in the Y3 \Balrog analysis. The \texttt{avg\_bdf\_mag} selection removes objects too faint to be detected in the DES Y6 sample. The quantity is computed by averaging the fluxes in the $riz$ bands, and then converting this average flux to a magnitude. The selection of $\texttt{avg\_bdf\_mag} < 25.4$ follows the Y3 \Balrog analysis, where the chosen threshold is the magnitude of injected sources that are detected with 1\% efficiency \citepalias{BalrogY3}. We have used the same choice in Y6 \Balrog, even though the Y6 data has a fainter magnitude limit than the Y3 data, to further optimize the number of usable injections from \Balrog Y6 relative to Y3. The selection criteria for the source and lens galaxy samples are at least one magnitude brighter than this limit (see histograms in Figure \ref{fig:StarGalSep}). Finally, the \texttt{bdf\_T} selection removes large objects in order to reduce the blending of injected, synthetic sources with other synthetic sources. These selections follow exactly those used in \citetalias{BalrogY3} for Y3 \Balrog.

For a subset of injections in Y6 \Balrog, we apply a weighting to preferentially inject galaxies from the deep field rather than probable stars.\footnote{We use the terminology of ``probable stars'' to refer to the objects in the DES Y3 deep fields \citepalias{DES_DF_Y3} that were classified as stars using photometric colors (and \textit{not} morphology).} While \Balrog will still contain injections of stars found in the deep field (as only a subset, and \textit{not all}, of the injections are weighted in this manner), we reduce the ratio of stellar injections to galaxy injections by performing this selection, as the latter are the sole focus of all the cosmology analyses. In practice this weighting down-weights objects in an approximately rectangular region in the size-magnitude plane. These star weights are defined as,
\begin{align}\label{eqn:star_weights}
w(m, T) = 1 - & \,\, \bigg[\bigg(0.9995 - \frac{0.9995}{1 + \exp[-\frac{T-0.2}{0.1}]}\bigg) \nonumber\\
& \times \bigg(0.9995 - \frac{0.9995}{1 + \exp[-\frac{m-22}{0.5}]}\bigg)\bigg]
\end{align}
using $i$-band magnitude, $m$, and size $T$ in units of arcsec$^2$. The weighting above removes objects with $\texttt{size} \lesssim (0.2\arcsec)^2$ and $\texttt{mag\_i} \lesssim 22$. Figure \ref{fig:weightscheme} shows the region of magnitude-size space that is down-weighted by this procedure. See Section \ref{sec:InjWgtscheme} below for more details on the subsamples.

\subsection{Injection and weighting scheme}\label{sec:InjWgtscheme}

\begin{figure}
    \centering
    \includegraphics[width = \columnwidth]{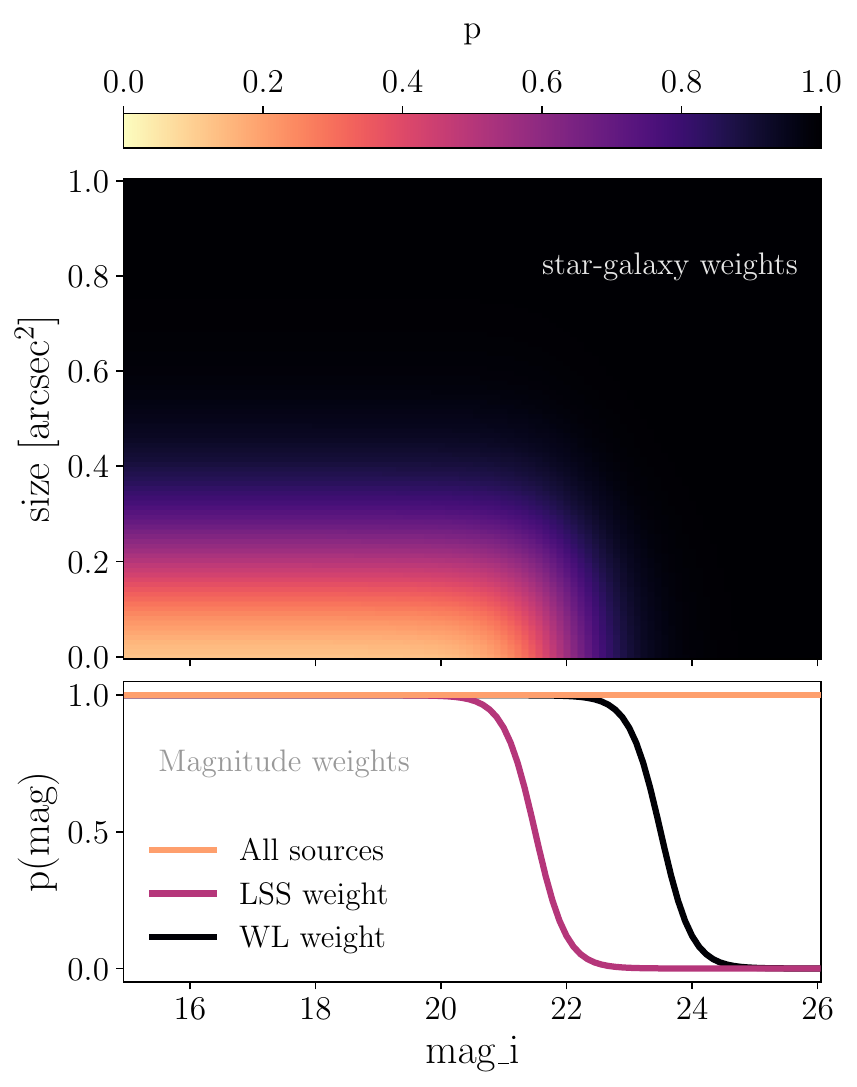}
    \caption{\textbf{Top:} the star/galaxy weights of the deep-field objects as a function of their size and magnitude. The weighting reduces the number of small and bright objects that are injected, which are predominantly stars. \textbf{Bottom:} The magnitude-based weights for the LSS and WL samples we inject into \Balrog. They are both sigmoid functions, with the transition centered at 21.5 and 23.5 in $i$ band for the LSS and WL sample, respectively.}
    \label{fig:weightscheme}
\end{figure}

\Balrog runs are performed for a single coadd tile at a time. For a given tile/run, the injections are assigned to a regular grid spanning the $0.7 \times 0.7 \deg^2$ region of the sky covered by the coadd tile (with a padding of $10 \arcsec$ from the edge). Once the sky location is defined, the source is injected into the relevant subsection of each CCD image that covers that region of the sky. Similar to DES Y3, the sky locations are a hexagonal grid (in pixel coordinates) with a separation of $20\arcsec$. This separation is chosen as we only inject sources with half light radii below $10\arcsec$ (see Equation \ref{eqn:DFSampleCuts}). Once the locations are defined, we randomly draw objects from the sample defined above in Section \ref{sec:SourceCat}. A new choice of note in Y6 \Balrog is separating our source injections into five categories, in order to boost statistics for the objects that are most important for downstream DES Y6 cosmology analyses. In each coadd tile, we inject sources belonging to each of the five categories:

\begin{itemize}
    \item[1.] \textbf{All sources (Y3-scheme):} A third of all injections in a given tile are of objects defined in Section \ref{sec:SourceCat} above, through Equation \eqref{eqn:DFSampleCuts}. Note that this includes both galaxies \textit{and stars}, i.e. we have not applied any star-galaxy selections in this sample.

    \item[2.] \textbf{LSS sample:} A sixth of the injections are a weighted sampling of the objects. These down-weight all galaxies with magnitudes fainter than $m_{i, {\rm ref}} = 21.5$, and also include the star-galaxy selection (see Figure \ref{fig:weightscheme}). The magnitude weights are defined as,
    \begin{equation}\label{eqn:mag_weights}
        w(m, m_{\rm ref}) = 1 - \frac{1}{1 + \exp[-4(m - m_{\rm ref})]}
    \end{equation}
    and we apply them using the de-reddened $i$-band magnitudes of the deep-field objects.

    \item[3.] \textbf{LSS high-quality redshifts sample:} Another sixth of the injections are the same as the LSS sample described above, but now with the additional requirement that the deep-field object being injected also has a redshift estimate from \textsc{Cosmos} \citep{COSMOS_2016, Weaver:2022:Cosmos}, \textsc{Paus} \citep{PAUSCOSMOS_2021},  or \textsc{C3R2} \citep{Masters:2017:C3R2_DR1, Masters:2019:C3R2_DR2, Stanford:2021:C3R2_DR3}. These objects are more vital in the redshift estimation method used in DES Y3 \citep{Myles:2021:SOMPZ, Giannini:2024:DES} and Y6 \citep{Y6SOMPZmdet, Y6SOMPZBFD, Y6SOMPZMaglim}.

    \item[4.] \textbf{WL sample:} Another sixth is once again the same as the LSS sample (including the star-galaxy selections), but with the magnitude-based weights moving to higher magnitudes as the weak lensing sample is fainter; here, we down-weight all objects fainter than $m_{i, {\rm ref}} = 23.5$.

    \item[5.] \textbf{WL high-quality redshifts sample:} The final sixth is similar to the LSS high-quality redshifts sample, but now tuned to the weak lensing sample by moving the magnitude-based weights to higher magnitude as mentioned above.
\end{itemize}

Figure \ref{fig:weightscheme} presents the different weights we use. The top panel shows the star-galaxy weights used to remove the stellar locus, and the bottom panel shows the different magnitude-based weights. The magnitude weights are sigmoid functions centered at values detailed above. This weighting scheme greatly increases the number of usable injections --- defined as injections that are actually detected in a given sample/method --- in Y6 \Balrog and thereby improves the statistical power of all analyses using \Balrog. The number of selected sources in Y6 \Balrog is increased, relative to what would be obtained using the ``Y3-scheme'' alone, by factors of 12, 2.7 and 1.6 for the LSS sample (\maglim, \redmagic), WL sample (\mdet, BFD), and \fitvd catalog (with no selections on magnitude and color) respectively. All estimates come from comparing the number counts of selected/detected objects from the ``Y3-scheme'' to those from the other four weighting schemes (while accounting for differences in the number of injections assigned under each scheme).

We note that the magnitude cutoffs defined above for the LSS and WL samples are slightly brighter relative to the actual magnitude distributions of the real samples \citep{Y6Mag, Y6Mdet, Y6BFD}, as our cutoff choice was defined prior to the availability of the Y6 sample definitions. However, we stress that the \Balrog synthetic catalog --- given the unweighted ``Y3-scheme'' subset of the catalog --- still contains objects across the full magnitude range relevant for all the Y6 cosmology samples. Any difference in our chosen magnitude cutoff and the true magnitude distribution of the Y6 samples will only result in a potential sub-optimality (i.e. increased noise) of measurements derived from \Balrog and \textit{not} in biases in these measurements.

\subsection{Fiducial run and magnification run} \label{sec:FidAndMag}

In Y3 \Balrog, a number of different variations were run, each serving a slightly different role in the main DES analysis \citepalias[][see their Table 1]{BalrogY3}. In this work, we perform two variations: a fiducial run, which injects the sources described in Section \ref{sec:SourceCat}, and a magnification run, where we inject the same sources but with their fluxes and area increased by 2\%. Note that all random seeds (during both the injection step and the subsequent postprocessing of images to catalogs) are fixed between the two runs. The magnification run is necessary for the Y6 cosmology analysis, as it enables estimates of the galaxy magnification coefficients for the specific samples used in DES Y6; for example, see \citet{Magnification_Y3} for how \Balrog was used to do this in DES Y3, and also Section \ref{sec:sec:mag} below. These magnification coefficients are folded into both the cosmological inference of galaxy clustering \citep{Pandey:2022:Redmagic, Prat:2022:GGL_Y3, DES:2022:3x2pt_Y3} and the redshift calibration analysis \citep{GattiGiannini:2022:Wz, Giannini:2024:DES, Y6WZ}.

\subsection{Improvements relative to \Balrog in DES Year 3}\label{sec:Y3toY6}

We start with a brief note on the connection between \Balrog and the image simulations \citep{Y6Imsims} used in DES Y6.
The purpose of \Balrog is to estimate the transfer function in as realistic a setting as possible. This is enabled by directly using the real CCD images. The produced synthetic catalogs then include the complexity of the imaging data and the distribution of image quality/properties across the survey. The purpose of the image simulations is to estimate any biases in the shear/redshift algorithms, by testing said algorithms on realistic models of our imaging data where we know the true properties for every source in the image. The image simulations, relative to \Balrog, are not as ideal a representation of the data and are therefore not used for estimating the transfer function. The \Balrog approach, relative to that of the image simulations, does not have the true properties of all sources in the images (namely, this information is missing for the real sources that already exist in the image) and so cannot be used in lieu of the image simulations. We have highlighted the broad, qualitative differences in approaches, which is then followed by differences in the exact technical implementations of the synthetic source injection pipelines (we do not detail this below for brevity).

\textbf{Adopting the \textsc{Eastlake} pipeline:} A key (logistical) difference from Y3 is the unification of our \Balrog pipeline with that used for the image simulations \citep{Y6Imsims}. Both \Balrog and the image simulations require source injection, though with some significant differences in final analysis choices: for example, the image simulations generate fully simulated images, by adding together the synthetic galaxies with simulated noise, star  sources, data-based background and mask etc. Whereas in \Balrog we inject galaxies (and probable stars) directly into the real images and therefore do not need to simulate any additional image properties. However, the vast majority of the processing steps are common between the two pipelines, and thus in DES Y6 both pipelines derive from the \textsc{Eastlake}\footnote{\url{https://github.com/des-science/eastlake}} codebase which performs most of the image processing steps discussed in Section \ref{sec:DESDM_Pipeline}. The specific \Balrog injection (Step 3) is performed outside \textsc{Eastlake}, as are the BFD measurements (Step 9) and matching (Step 10). All remaining steps associated with image processing (Step 1 to 8) are performed using \textsc{Eastlake} and the packages within, and are therefore \textit{identical} between this work and the image simulations used for shear calibration in DES Y6 \citep{Y6Imsims}.\vspace{10pt}

\textbf{Moving to \textsc{Piff} PSF models:} In Y6 \Balrog, our multi-epoch multi-band object fits use the \textsc{Piff} models \citep{PIFF_Y3} rather than following Y3 \Balrog in using the \textsc{PSFex} models \citep{Bertin2013}. This improves the consistency of our synthetic catalog for the two source samples, \mdet and BFD, and leaves a minor mismatch with the \fitvd catalog in Y6 \citep{Y6Gold} as the latter is constructed using \textsc{PSFex} models. \vspace{10pt}

\textbf{Identification of blended objects:} In Y3 \Balrog, the matching step included an additional component --- beyond that described in step 10 of Section \ref{sec:DESDM_Pipeline} --- to identify \Balrog injections that were blended with a real source in the image. In this step, an injected source was considered problematic only if it was blended \textit{and} it was the fainter object in the blended set. In DES Y6, we identify all detected synthetic injections that have a real object within $1.5\arcsec$ of their location. All such synthetic objects are flagged in our catalogs, regardless of whether or not they are the brighter source in the blended group. This is a more conservative selection relative to DES Y3. Such objects constitute only $<0.2\%$ of the entire catalog, and are removed from all analyses that use \Balrog, including all those presented below.\vspace{10pt}

\textbf{Weight scheme and star catalog:} As described above, the weighting scheme used in Y6 was not part of the DES Y3 analysis. Under this new scheme, a subset of the injections (1/3rd of all injections) consists of any real, detected object (i.e., objects that are not artifacts) in the deep fields. This includes stars. Thus, in Y6 our star injection sample uses real stars rather than a simulated star sample as was done in Y3. At brighter magnitudes this enhances the realism of the recovered star sample as we are directly injecting stars from the deep field. However, at fainter magnitudes this means we rely on the accuracy of the data-driven, color-based star-galaxy classifier of \citetalias{DES_DF_Y3} to determine whether a given deep-field object (which is then injected into images via \Balrog) is a faint star or a faint galaxy. \citetalias{DES_DF_Y3} (see their Figure 15) show that the star sample is $>95\%$ pure down to $\texttt{mag\_i} < 24$. All subtleties about the star-galaxy classifier are relevant only when using \Balrog for analyses of stars. Additionally the color/magnitude distribution of the injected stars will, by construction, follow that found in the Y3 deepfields whereas in reality this distribution varies across the DES footprint. \vspace{10pt}

\textbf{\Balrog for the full DES footprint:} In DES Y3, the \Balrog sample was restricted to around $2500$ tiles due to computational limitations. In DES Y6, this is increased 5 fold, with over $12000$ tiles being run: $10,169$ tiles for the fiducial run, and $2,000$ for the magnification run. The full coverage of the survey enables new science cases, particularly any area-dependent studies of all \Balrog-related science, such as magnification, transfer functions, imaging systematics, etc. Figure \ref{fig:footprint} shows the area covered by \Balrog in comparison to the DES Y6 data. There is already visual indication that \Balrog traces the large scale density fluctuations across the survey (due to variations in depth and other observing conditions);\footnote{The amplitude of fluctuations is suppressed because the weighted injection scheme in Y6 Balrog (see Section \ref{sec:InjWgtscheme}) results in the majority of injected objects being above the object detection threshold. The fluctuation amplitude in Figure \ref{fig:footprint} will increase if we limit the sample to the unweighted injections alone.} for example, the galaxy number density in both \Balrog and data is lower near the east and west (left and right) edges of the footprint, which are regions of high stellar density at lower Galactic latitude. These regions have a lower effective area due to masking around bright stars, and are regions where objects are more frequently observed as blended sources. Other correlations with observing conditions are noted below in Section \ref{sec:sec:sec:sysmaps}. Note that these large-scale fluctuations in the number density of \Balrog objects are not generated by hand, and manifest solely from the interaction of image properties (such as noise, sky background, source crowding, etc.) and the object detection/measurement algorithms.

\section{Validation of the synthetic sample}\label{sec:photo}

As mentioned previously (and detailed in Section \ref{sec:applications}), the \Balrog data products are used extensively to meet many different DES science requirements. In this section, we highlight both the quality of the \Balrog catalog --- checking its consistency with data --- to justify its use in the DES Y6 cosmology pipeline, and also the science performance of algorithms involved in going from raw images to robust object catalogs. We first highlight in Section \ref{sec:sec:Consistency} the consistency of our catalog with DES data, then in Section \ref{sec:sec:performancefid} and \ref{sec:sec:performancestar} the performance of the photometric pipeline for the galaxy sample and star sample, respectively. Finally in Section \ref{sec:sec:stargalsep}, we detail the performance of the different star-galaxy classifications used in defining the various samples used in the DES Y6 cosmology analyses.

\subsection{Consistency with DES Data}\label{sec:sec:Consistency}

\subsubsection{Completeness}\label{sec:sec:sec:completeness}

\begin{figure}
    \centering
    \includegraphics[width = \columnwidth]{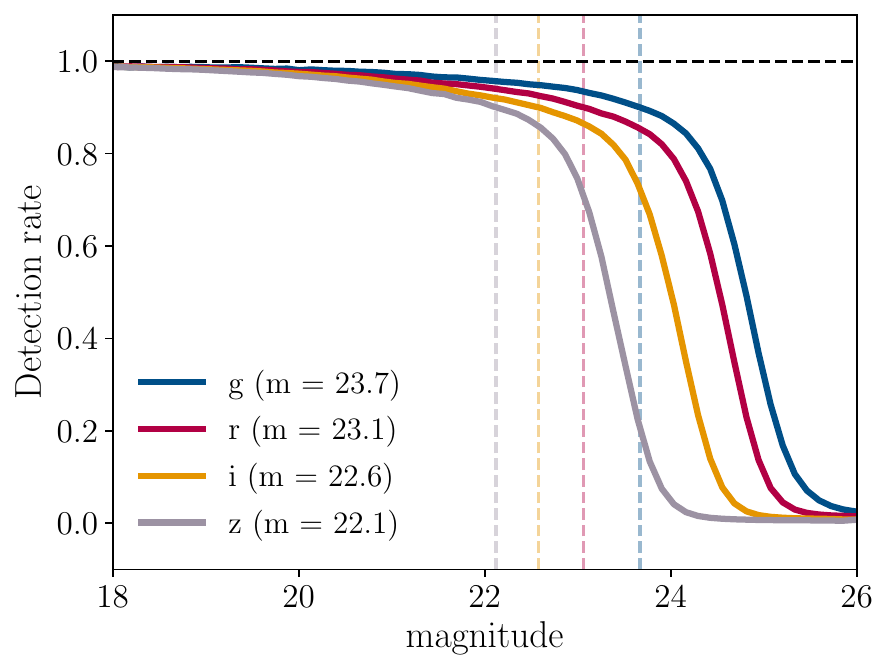}
    \caption{The detection rate of objects as a function of their magnitude. Objects are detected if they have $\texttt{SNR} = \texttt{flux}/\texttt{flux\_err} > 5$ in order to avoid any spurious detections. Vertical lines denote the magnitudes at which we achieve $90\%$ completeness, and the numerical values are given in the legend. This rate is measured using \Balrog objects across the full Y6 footprint.}
    \label{fig:DetectionRate}
\end{figure}

Figure \ref{fig:DetectionRate} shows the detection probability of a galaxy, as a function of its $griz$ magnitudes. As expected, the rate nears 1 for bright objects and drops to 0 for faint objects. The vertical dashed lines show the magnitude where we have $90\%$ completeness of objects with $\texttt{SNR} = \texttt{flux}/\texttt{flux\_err} > 5$. We impose a signal-to-noise selection in each band to remove any spurious false-positive \Balrog detections due to matching injected sources with noise fluctuations; this is relevant only near the faintest end ($m \geq 25.5$) and particularly for the $g$-band as it is not used in the detection step. When considering all objects with magnitudes brighter than the threshold set by the $90\%$ completeness rate, more than $99.9\%$ of the sample have $\texttt{flux}/\texttt{flux\_err} > 5$. For objects at the $90\%$ completeness rate we find $\langle \texttt{flux}/\texttt{flux\_err} \rangle \approx 20$.

While our test of completeness does not include direct comparison to the Y6 data --- unlike in the Y3 analysis, where a data-based estimate of completeness was done by matching the deep-field detections to their wide-field counterparts --- this measurement shows us that \Balrog produces physically reasonable completeness rates for the extreme ends of bright and faint objects, which in itself is a test of consistency. Estimates of the improvement in depth from DES Y3 to  DES Y6 are presented alongside the Y6 Gold Catalog \citep{Y6Gold} --- and the DES DR2 \citep{DES_DR2} --- and we find consistent improvements in \Balrog. Section \ref{sec:sec:detectioneff} below details one such comparison.

\subsubsection{Photometry and morphology}\label{sec:sec:sec:photomorpho}

\begin{figure*}
    \centering
    \includegraphics[width = 1.5\columnwidth]{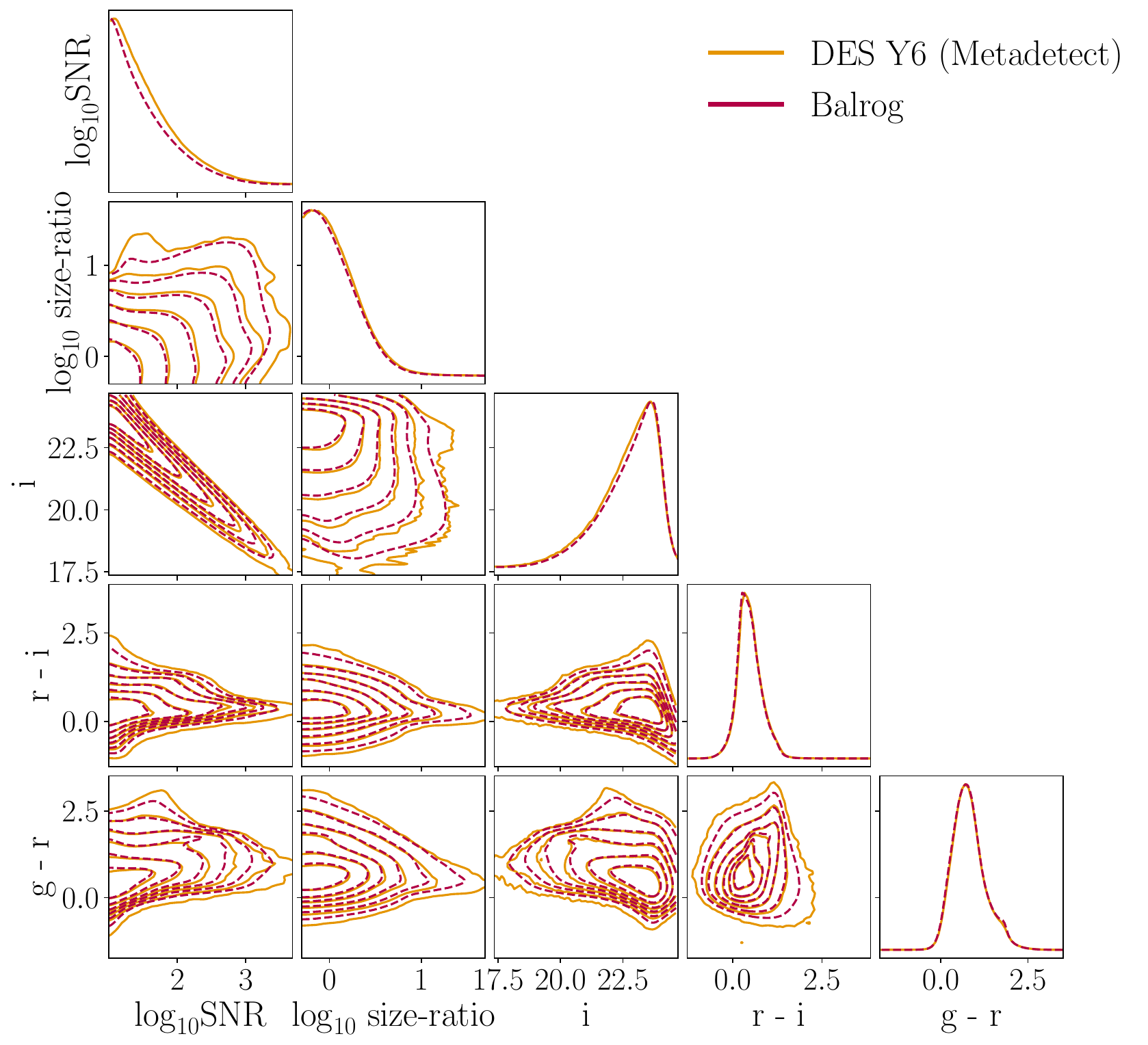}
    \caption{A corner plot of the distributions of galaxy properties from the \mdet catalog. The contours show the $0.5\sigma$ to $3\sigma$ bounds, in increments of $0.5$. The Balrog density contours agree well with the real data except for the rarest (3$\sigma$) sources' distribution. Thus, nearly 99\% of the catalog is consistent between \Balrog and data. The properties shown are the signal-to-noise, the ratio of object size and PSF size, the $i$-band magnitude, and the $r - i$ and $g - r$ colors.}
    \label{fig:CompareMdet}
\end{figure*}

A key consistency check is ensuring the distribution of measured properties in a given \Balrog catalog matches that of the corresponding data catalog. Figure \ref{fig:CompareMdet} shows a range of properties of the \mdet sample from \Balrog and data; we have shown only properties for the unsheared sample here, but the remaining four samples show similar consistency as the unsheared sample. In general, the data and \Balrog distributions match across $\approx 99\%$ of the data volume. Note that we expect some difference between the contours, as an implicit assumption in \Balrog is that the deep-field galaxies are a representative sample of the wide-field galaxies, whereas in reality, simple factors like sample variance (arising from the small footprint of the deep fields) will break this assumption slightly. Figure \ref{fig:CompareMdet} shows such factors are likely a minor contributor given that the data and \Balrog catalogs match for 99\% of the data volume. The distributions are similarly consistent when repeating this check for other galaxy samples (BFD, \textsc{Maglim++}, and \textsc{RedMaGiC}). We do not show those comparisons here for brevity.

Note that the above measurements are performed using only the unweighted subsample in \Balrog (the ``Y3-scheme'' defined in Section \ref{sec:InjWgtscheme}). Adding morphology/photometry-based weighting to the deep-field sample during source injection will generate large differences between the injected sample and the actual data. For example, under the LSS weights any galaxy with $m_i \gtrsim 23$ is never injected, even though the weak lensing sample contains objects down to $m_i \sim 24.5$. Thus, comparing the distribution of properties in data while using the weighted subsample in \Balrog will naturally lead to discrepancies in the comparisons. While this could be handled by ``inverse weighting'' the different, weighted distributions, we opt for the simpler approach of just using the unweighted ``Y3-scheme'' sub-sample.

\subsubsection{Spatial Variation and Property Maps}\label{sec:sec:sec:sysmaps}

\begin{figure*}
\includegraphics[width = 2\columnwidth]{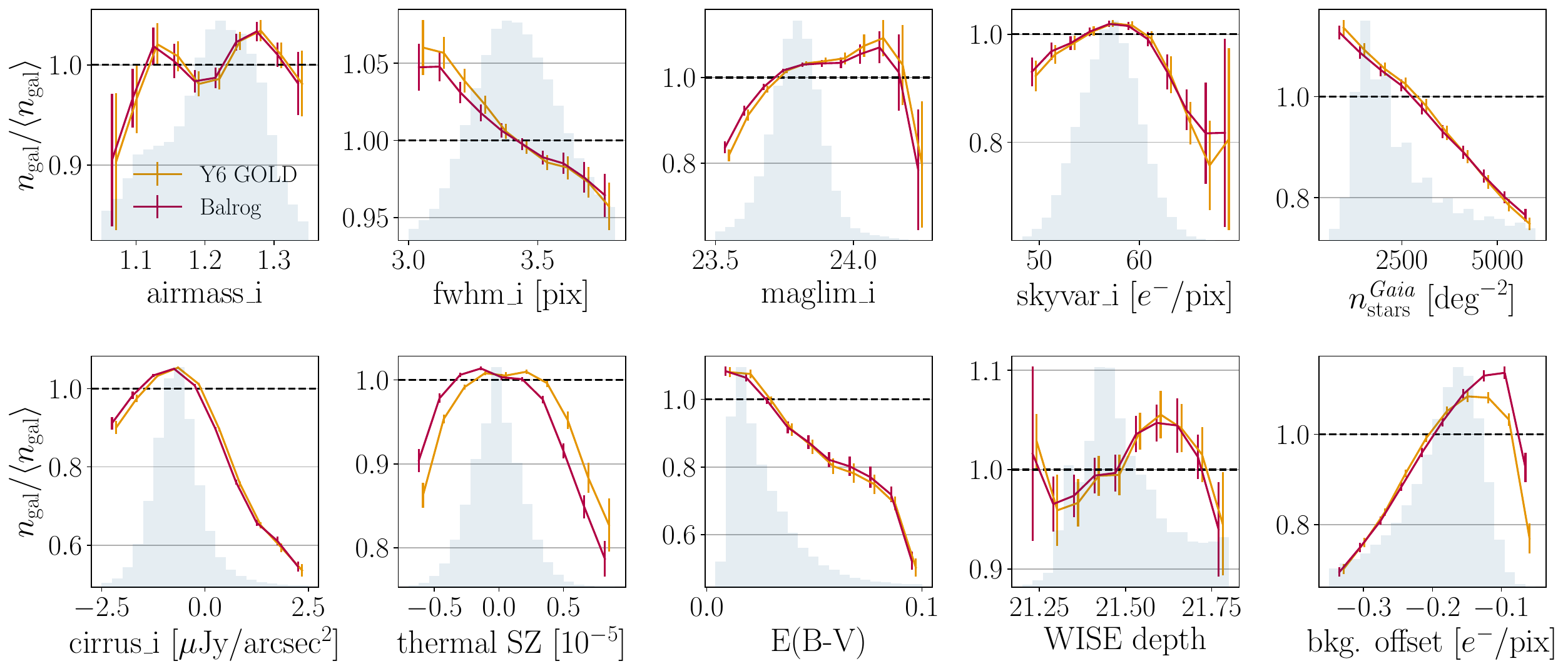}
\caption{The trend in number density fluctuations $n_{\rm gal}/\langle n_{\rm gal}\rangle$ as a function of various survey observing properties, for the Balrog (red) and Y6 Gold (orange) samples for overlapping HEALPix pixels of \texttt{NSIDE=512}. We apply basic selections, such as star-galaxy classifications, to both samples. The distribution of survey property values for the Y6 Gold map is shown in the background (blue gray) of each plot to highlight typical values. The errors have been estimated using a jackknife resampling of the pixels with 100 patches. We select only a few properties to show in this figure and, for those maps with different values on the \textit{griz} photometric bands, we only show the \textit{i}-band. Briefly defined, these survey properties are: the airmass, the PSF size, the magnitude limit at $S/N = 10$, the sky brightness variation, the \Gaia stellar density, the Galactic cirrus nebulosity, the Sunyaev Zel'dovich effect, the interstellar extinction, the WISE depth, and the background offset. The $\chi^2$ between the data and the \Balrog measurements follows $\chi^2 / N_{\rm dof} \lesssim 1$ for all cases except the correlations with the thermal SZ and the background offset, where $\chi^2 / N_{\rm dof} \gg 1$. This is expected, due to the impact of large-scale structure on these maps and we detail this in Section \ref{sec:sec:sec:sysmaps}.}
\label{fig:survey_properties}
\end{figure*}

The detection and characterization of an object on the sky will depend on the observing conditions of the processed images, and on other non-cosmological fields observed in the images, such as Galactic dust or stellar density \citep{Ho_2012, Ross_2012, Leistedt_SDSS_2013, Rykoff:2015:Maglimits}. We represent variations of these effects across the DES footprint with ``survey property'' (SP) maps \citep{Y6Gold, Y6Masking}. The \Balrog catalogs will be a more accurate representation of the data if they also capture the dependence of object properties on such survey property maps. This correlation of galaxy properties with observing conditions is a key source of systematic uncertainty to be addressed for measurements of the cosmological clustering of galaxies \citep{Monroy:2022:DESY3, Kong:2024:ObiWan}, and thus reproducing it in \Balrog is of importance.

We compute the correlation between galaxy number counts and survey properties in \Balrog and in the Y6 Gold sample, and compare the two. At zeroth order, survey property maps alter the measured properties of galaxies in our images, and this will lead to variations in the number of detected/selected galaxies in different regions of the sky. For example, regions of high airmass have larger PSFs and can lead to fewer faint/small galaxies being observed, which in turn will reduce the overall galaxy number density in such regions. We measure these galaxy--SP map correlations by making maps of the galaxy number counts, $N_{\rm gal}$, and measuring the average $N_{\rm gal}$ as a function of SP map value. We use \textsc{HealPix} \citep{Healpixels} maps of $\texttt{NSIDE} = 512$, but note that other choices result in similar agreement between \Balrog and data.

For a given SP map, we assign healpixels into 20 equal-size bins, spanning the 1-99 percentile range of the map values in order to avoid outliers. Within each of these 20 bins, we compute the mean galaxy number density. This process is repeated for both the \Balrog and Y6 Gold galaxy samples. We use the \fitvd catalogs, with $\texttt{EXT\_MASH} = 4$, so that our sample only includes galaxies that could be part of the lens sample used in the cosmology analysis. In Appendix \ref{appx:SPmaps_Lens}, we also show results from repeating this test but now using the full selection function of the \maglim and \redmagic samples. The uncertainty on the measurements is estimated by splitting the DES footprint into 100 patches and performing a leave-one-out jackknife resampling.

We use SP maps spanning survey observing conditions, maps of Galactic properties, and maps of other large-scale structure probes. In particular, our observing conditions include the airmass, full-width half-max of the PSF, the magnitude limit (or depth), the variation in the sky background, and the background offset. These are measured using \textsc{Decasu}\footnote{\url{https://github.com/erykoff/decasu}} which uses properties of individual CCD images to make sky maps of observing conditions for the survey/dataset. We only present the correlations with the $i$-band maps but our results and conclusions are the same for the other bands as well. We then use the depth map in the \textsc{WISE} survey \citep{Wright:2010:WISE}, as the \textsc{Maglim++} sample is defined using photometry from the \textsc{unWISE} program \citep{Meisner:2017:Unwise}, which includes data from the \textsc{WISE} survey. We then use the Galactic stellar density, estimated through \Gaia \citep{Gaia_DR3_2022}, the Galactic cirrus \citep{Y6Gold}, and the extinction $E(B - V)$ map from \citet{Schlegel_1998}. Finally, we also include the Sunyaev-Zeldovich (SZ) map from the Atacama Cosmology Telescope \citep[ACT, ][]{ACT:2007, ACT:2016}, which is a tracer of the large-scale structure, and contains emissions from galaxies over a wide range of redshifts \citep[see][for a review of the SZ effect]{Carlstrom2002SZReview}. This map is not used as a systematics map but rather as a tracer of large-scale structure (not measured in DES) for validating the systematics correction of the lens samples; see \citet{Y6Lens} for more details. We note that the background offset map mentioned prior is also a potential tracer of large-scale structure as the excess presence of (unresolved) galaxies can increase the mean sky background and lead to larger background offsets.

Together, these maps span a wide variety of quantities that can impact measurements of galaxy properties (or correlate with them) in different ways. Figure \ref{fig:survey_properties} shows the mean galaxy number density as a function of SP map value. The background of each panel (light gray histogram) shows the distribution of survey property values across the DES Y6 footprint. The galaxy--SP correlations in \Balrog are statistically consistent ($\chi^2 / N_{\rm dof} \lesssim 1$) with those present in the data, indicating that \Balrog accurately captures the correlations between galaxy properties and observing conditions.

The two exceptions are the correlations with the ACT SZ map and the background offset map. However, these two exceptions are in fact expected. Both maps are tracers of large-scale structure as mentioned above. The correlations between the Y6 data and these maps will therefore contain a \textit{cosmological} correlation. Such a cosmological term is not present in \Balrog by construction as the synthetic galaxies are distributed uniformly across the coadd tile. Thus, there will be an expected difference in the galaxy--SP correlations between \Balrog and the Y6 data for these maps. 

There is also a correlation between the galaxy counts and the \textsc{WISE} depth (for both \Balrog and data). The galaxy sample here does not use \textsc{WISE} data for its selection, and therefore any correlation observed is a consequence of the \textsc{WISE} depth being correlated with other observing conditions that \textit{do} impact the galaxy sample used in this analysis.

We note that \citet[][see their Figure 12 and 13]{Kong:2024:ObiWan}, who also perform SSI in DECam images --- but to study the targets selected for the DESI survey --- find significant differences in the correlations of SSI galaxies with the extinction map compared to that of real galaxies with this map. However, these differences are prominent in the northern Galactic hemisphere and not present in the southern hemisphere, where the DES footprint is located. They suggest this qualitative difference across areas is also related to how their imaging data in the northern half was observed using a depth-based strategy that could lead to residual impact from extinction corrections (see their Section 7.2). The same is not present in the southern half, as that region is dominated by DES data which does not follow this observing strategy. They also suggest this difference between the north/south regions highlights systematics in the northern regions of the dust extinction map of \citet{Schlegel_1998}. As shown above, our work does not find any significant discrepancies in the correlations of \Balrog galaxies with the extinction map. This is also true for the individual lens samples, which we detail in Appendix \ref{appx:SPmaps_Lens}.

\subsection{Photometric Performance of Galaxy sample}\label{sec:sec:performancefid}

\begin{figure*}
\includegraphics[width = 2\columnwidth]{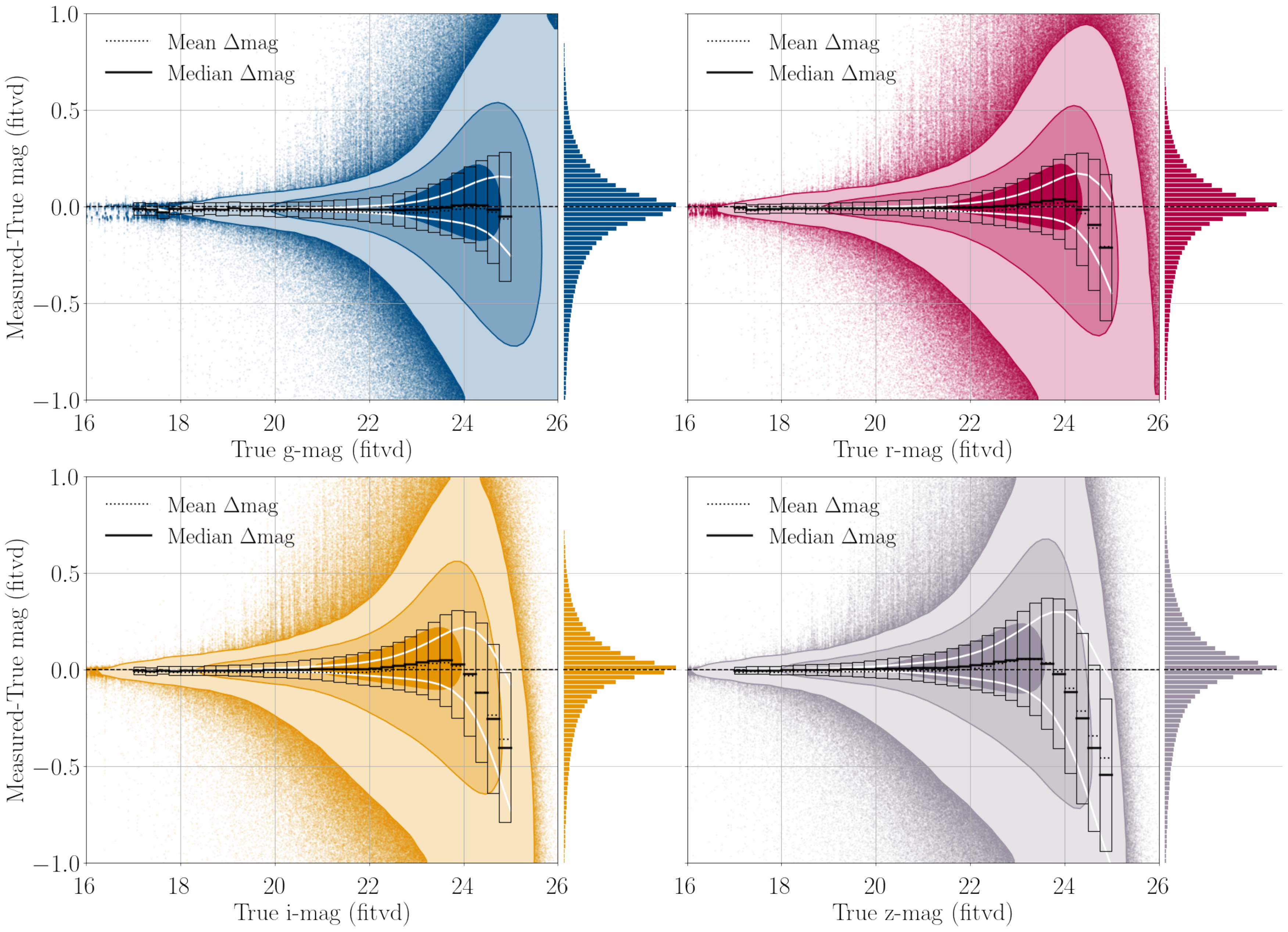}
\caption{The distribution of differences in recovered \fitvd \textit{griz} magnitude vs the injected magnitude ($\Delta\mathrm{mag}_{\mathrm{DF}}$) as a function of input magnitude for the galaxy sample. The contour lines correspond to areas containing 39.2\%, 86.5\%, and 98.9\% of the data volume, which are the $1,2,3\sigma$ bounds of a 2D Gaussian, respectively. The mean (dotted), median (solid), and standard deviation (bars) of the magnitude responses, in bins of size 0.25 magnitude, are also shown. The white lines are the average magnitude error of objects in a given bin, where the error is estimated during \fitvd fitting. There is some discretization in true magnitude (particularly in the brighter end) given we have only a small number of bright objects in the deep-field sample.}
\label{fig:y6deepmags}
\end{figure*}

The DESDM processing pipeline is multi-faceted, and includes many different steps (each validated and tested) that are integrated to perform the entire image processing procedure; see \citet{DESDM_ImageProcessingPipeline} for details of the pipeline used in DES, and \citet{DES_DR2, Y6Gold} for modifications made specifically in DES Y6. \Balrog provides a precise test of the pipeline's performance by comparing the properties of injected objects to their measured properties in noisy images. We perform this test with the Y6 dataset, using the \fitvd measurements of the galaxies. The deep-field galaxies also have \fitvd measurements made using the same algorithm, and therefore can be used to precisely quantify any biases in the wide-field measurements.

Figure \ref{fig:y6deepmags} compares the differences between the measured and true deep-field magnitudes $\Delta\mathrm{mag}_\mathrm{DF}$ as a function of true magnitude in the \textit{griz} bands. The density contours include 39.3\%, 86.5\%, and 98.9\% of the population, which correspond to the 1,2, and 3$\sigma$ regions of a 2D Gaussian. The mean bias, median bias, and standard deviation of the differences are over-plotted as a dotted line, solid line, and black bars, respectively. We compute these in bins of 0.25 magnitudes. The standard deviation is computed using the 16\% and 84\% percentiles to avoid biases from outliers. The median and standard deviation of the magnitude differences are tabulated in Table \ref{tab:photo_acc}. The white lines for each also show the $1\sigma$ magnitude region --- as computed by the \fitvd algorithm --- around the median difference. Similar to the results of \Balrog in DES Y3 \citep[][see their Figure 16]{BalrogY3}, the estimated errors are smaller relative to the scatter in the measured magnitudes of the \Balrog synthetic sources (the black, vertical bars in Figure \ref{fig:y6deepmags}). The \Balrog-based estimate of this error is generally expected to be larger as it includes systematic effects like the variations across different observing conditions, which is not accounted for in the internal, \fitvd-based error estimates. The differences between the \Balrog-based and \fitvd-based error estimates are larger for brighter objects, where such systematic effects are the dominant contributor to the magnitude differences, and are smaller for fainter objects where noise fluctuations are the dominant contributor.

The tails to negative magnitude offsets in Figure \ref{fig:y6deepmags} was also found in \citetalias{BalrogY3} (see their Section 4.3.3), and can arise from a number of different causes, which include excessive pixel masking, residual light from neighboring sources etc. Their Figures 8 and 19 show the same asymmetric tails, at faint magnitudes, that we find in Figure \ref{fig:y6deepmags}. We also note that some negative bias (i.e. measured magnitudes are brighter than the true magnitudes, on average) is expected for faint objects due to the detection bias produced by applying a signal-to-noise selection during object detection. If we consider multiple wide-field realizations of a faint object, we preferentially detect realizations where the noise fluctuations boost the measured brightness. This is a natural consequence from using noisy data, and does not indicate problems with the image processing algorithms. A detection bias can also explain the relatively low bias in the $g$-band relative to the $riz$-bands, as the former is not used in object detection and is therefore akin to a forced photometry measurement, which will naturally avoid detection biases. Similarly the detection bias is expected to be stronger in Y6, relative to Y3, due to the lower signal-to-noise threshold. 

\begin{figure*}
\includegraphics[width = 2\columnwidth]{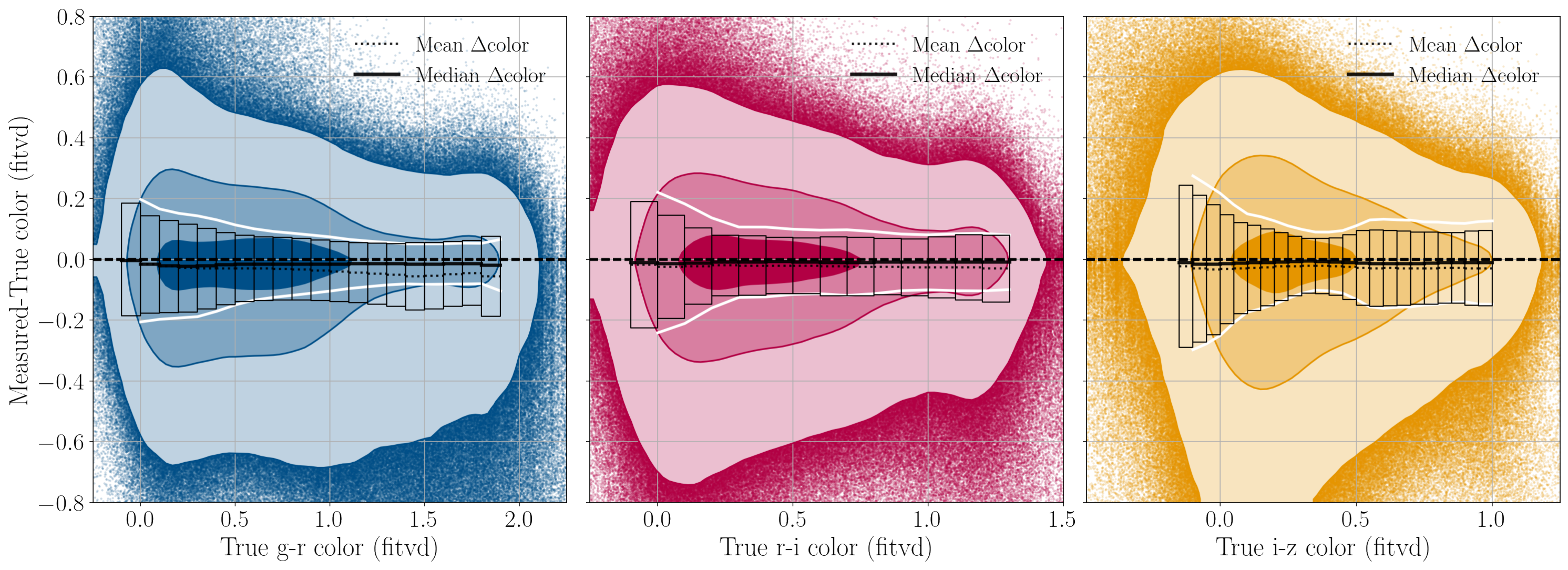}
\caption{Similar to Figure \ref{fig:y6deepmags} but for the differences in measured \fitvd $g-r$, $r-i$, and $i-z$ color. The statistics are computed in bins of 100 mmag magnitude for $g-r$ and $r-i$ and 50 mmag for $i-z$. The median bias in the recovered colors is below $0.02$. The \fitvd-based uncertainty on the colors, which is computed by adding the individual magnitude uncertainties in quadrature, is in good agreement with the \Balrog-based estimate.}
\label{fig:y6deepcolor}
\end{figure*}

\begin{figure*}
\includegraphics[width = 2\columnwidth]{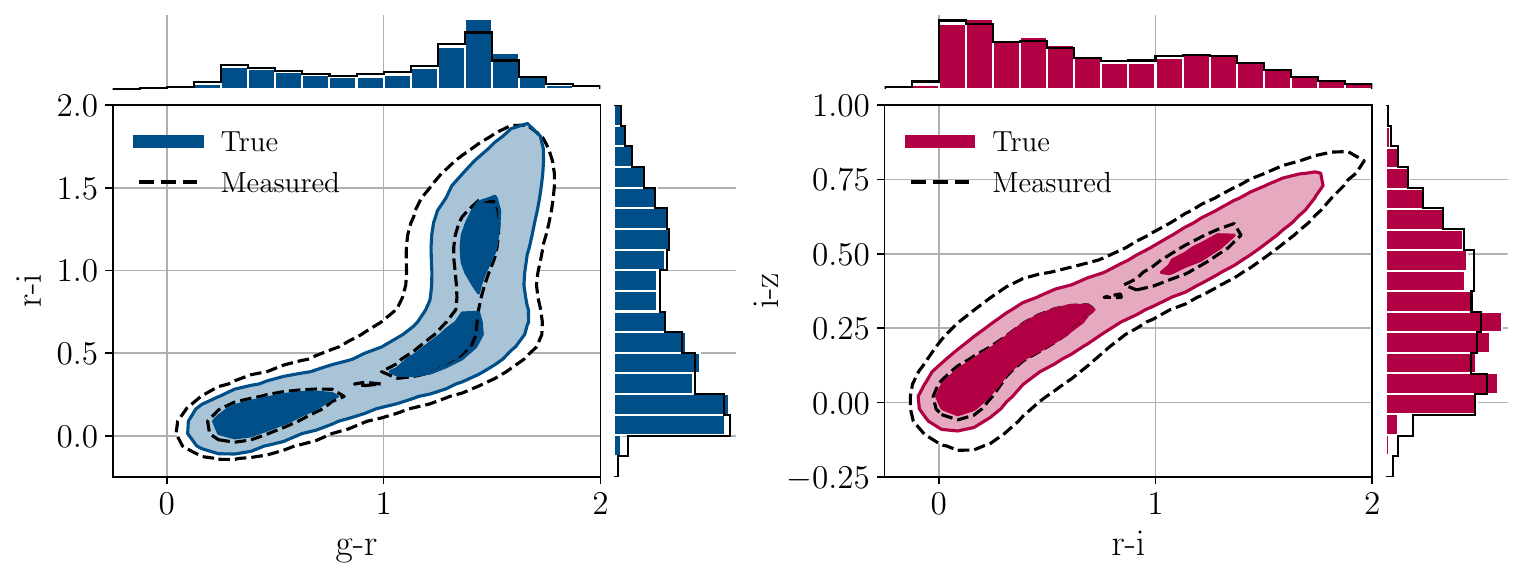}
\caption{The $g-r$ vs $r-i$ and $r-i$ vs $i-z$ color-color distributions of the stellar sample for the injected colors in blue and the measured colors in black. The density contour lines correspond to the percentiles of the first two sigmas of a 2D Gaussian, containing 39.2\%, 86.5\% of the data volume respectively. The marginal distributions are included for comparison.}
\label{fig:colorcolorstars}
\end{figure*}

The estimates in Figure \ref{fig:y6deepmags} are made using galaxies that pass all the Y6 Gold selection criteria as well as the star-galaxy selection, $\texttt{EXT\_MASH} \geq 4$ \citep{Y6Gold}. This is similar to, but still strictly distinct from, the galaxy samples used for the various cosmology analyses in DES Y6. Our estimates (see also Table \ref{tab:photo_acc}) show we recover the true magnitudes to within 3-6 millimag (below $m < 20$) for the $riz$ bands, and within 12 millimag for $g$-band. This is generally larger than, but the same order as, the relative photometric uniformity which is at $\lesssim 2$ millimag \citep{DES_DR2} and also smaller than the 11 mmag uncertainty on the absolute photometric calibration \citep{DES_DR2}. For fainter objects $(m > 23)$, we find larger biases of up to 50 millimag that increase to $\approx 500$ millimag at $m \sim 25$. We note that for $m < 24$, the median difference is always well below the standard-deviation. The differences we discuss here were also found in \citetalias{BalrogY3} (see their Figure 16), where they show that the bias grows increasingly negative for fainter galaxies. However, the Y6 dataset extends to fainter objects due to the increased depth, and also reduced the signal-to-noise threshold for detection from $S/N = 10$ to $S/N = 3$ \citep{Y6Gold}, relative to Y3 and we therefore observe stronger biases than was found in their work.

There is also a visible, positive bias of $\Delta m \sim 0.03$ in intermediate magnitudes, $23 < m_b < 24$. This was also found in the DES Y3 analysis of \citetalias{BalrogY3} (see their Section 4.3.1) and was identified to arise from errors in the sky-background estimates of the \SE algorithm, which are then propagated to the rest of the galaxy catalog processing. The \SE algorithm can overestimate the sky-background, often due to crowded fields. This erroneous background estimate, when subtracted from an image, results in measuring a fainter magnitude for a given detected object, and therefore a positive shift on the y-axis of Figure \ref{fig:y6deepmags}. This background-related bias grows in the presence of more crowded fields, and crowding increases towards the redder bands, resulting in the chromatic effect that we observe: the $g$-band has very little positive bias, but the $z$-band has a much more significant one. Note that this effect is present for \textit{all} objects but only becomes more apparent at fainter magnitudes, where an object's flux is more similar to the background level.

We also investigate the photometric accuracy of the object colors. Such colors are used extensively in applications such as photometric calibration, star-galaxy classification, redshift estimation, and studies of the Milky Way structure/dynamics \citep[\eg][]{FGCM, RedshiftMethods_2019, Myles:2021:SOMPZ, Giannini:2024:DES, DNF, Y6Gold, DNF, Shipp:Streams:2018, Pieres:2020:MW}. Figure \ref{fig:y6deepcolor} shows the difference in measured and true colors for $g-r$, $r-i$, and $i-z$. The density contours and over-plotted summary statistics are defined in the same way as in Figure \ref{fig:y6deepmags}. One additional selection is done to this sample: a signal-to-noise of greater than 5, where this signal-to-noise is directly estimated during the \fitvd fluxes and flux errors, to remove color measurements from particularly noisy detections. Our results show that the median difference in colors is consistently negative, suggesting a small bias for the recovered colors to be bluer than the true colors. However, this median difference is always within 10-20 millimag, and also similar across all three colors we present, indicating there are no band-dependent effects. This behavior is also found in \citetalias{BalrogY3} (see their Figure 18).

In conclusion, we find that the photometric accuracy of the DES Y6 dataset is similar to that of the Y3 dataset. The accuracy in recovering the true magnitudes degrades for fainter objects (particularly those objects that are too faint to be found in DES Y3 but are now detected in the deeper, DES Y6 data), as is expected. While we also discuss the faint regime, where the photometric recovery shows growing biases, all the main source and lens galaxy samples used in the DES Y6 cosmology analysis occupy the brighter end of the magnitude range in Figure \ref{fig:y6deepmags}, which show fluxes are recovered at the 2\% level.

\subsection{Photometric Performance of star sample} \label{sec:sec:performancestar}

The image processing pipeline --- in particular, the fitting of models to objects --- can show different accuracies when processing galaxies or stars, given the differences in morphology of the categories. We repeat the analyses above using only the stars injected into \Balrog; these are determined through the color-based classifier built and presented in \citetalias{DES_DF_Y3}. Their work (see their Figure 15) shows that the star sample is $>95\%$ pure down to $\texttt{mag\_i} < 24$.  For brevity we present the results in Appendix \ref{appx:performancestar} but summarize here that photometric measurements of stars are accurate to a similar level as seen in the galaxy sample (Section \ref{sec:sec:performancefid}). Table \ref{tab:photo_acc} lists the photometric accuracy for the galaxy and star samples, and shows that the median magnitude differences are similar for the two samples but the standard deviations of these differences is roughly two times smaller for the star sample.

In addition, Figure \ref{fig:colorcolorstars} shows the color-color diagrams for $g-r$ vs $r-i$, and $r-i$ vs $i-z$ colors, as measured in \fitvd, for both the true and measured star samples.  As expected, the measured colors have a broader distribution due to being noisy measures, but the non-linear shape of the 2D distribution is still preserved by the image processing pipeline.

\subsection{Contamination \& efficiency of the star-galaxy classification}\label{sec:sec:stargalsep}

\begin{figure*}
    \centering
    \includegraphics[width = 2\columnwidth]{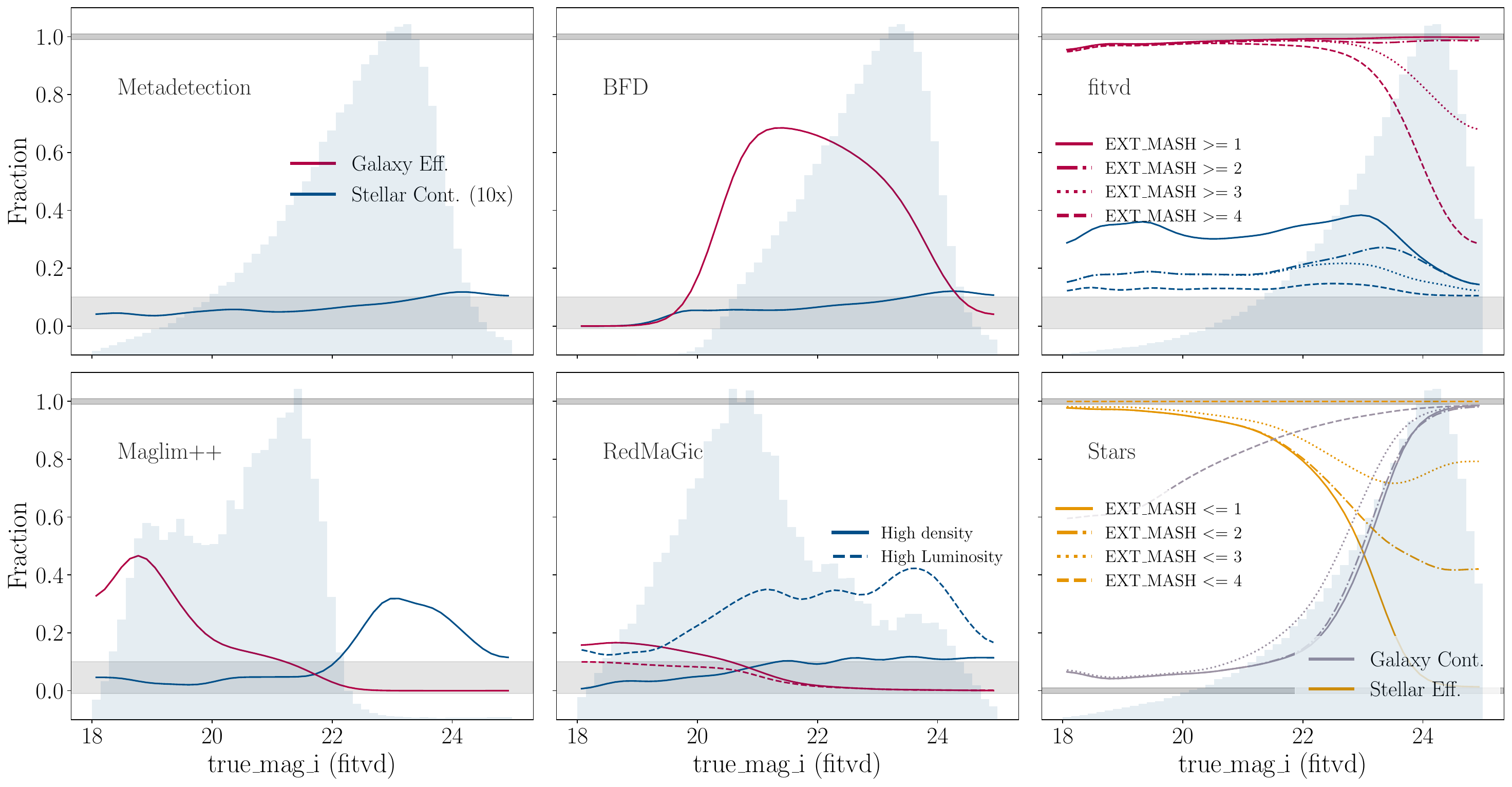}
    \caption{We show the efficiency (the number of selected good objects over the number of all good objects) and contamination (the number of selected bad objects over the total number of selected objects) for different samples used in DES Y6. We multiply the stellar contamination by 10x for easier visibility. The top row shows the two source samples --- \mdet and BFD --- and the base galaxy catalog from \fitvd. The bottom row shows the lens samples, from \maglim and \redmagic, while the right-most panel shows the efficiency and contamination for the \textit{star} sample based on the \fitvd catalog. The histograms in the background show the number counts of objects as a function of true, injected magnitude. For \mdet, we cannot compute a galaxy efficiency curve as the necessary data was not kept due to storage limitations. For \redmagic, we show the high density sample in the histogram, and for \fitvd and stars, we show the full sample without selections in the histogram. The gray bands show the $1\%$ region around fractions of 1 and 0, with the region around 0 being increased by 10x when corresponding to stellar (not galaxy) contamination curves.}
    \label{fig:StarGalSep}
\end{figure*}

The DES Y6 analyses contain many different galaxy samples: the \mdet and BFD source samples, and the \maglim and \redmagic lens samples. Through \Balrog we can quantify the contamination and efficiency of the star-galaxy classification as defined in the different sample selections. To do so, we estimate the ``galaxy efficiency'' defined as the fraction of true galaxies in the selected sample relative to \textit{all detected} true galaxies, and we estimate the ``stellar contamination'', defined as the fraction of all selected objects that are true stars. We classify \Balrog injections into galaxies and stars based on the color-based classifier of \citetalias{DES_DF_Y3}. The stellar contamination rate depends on the ratio of stars and galaxies in the truth catalogs. Therefore all our results below are valid for ratios of stars to galaxies similar to those found in the DES deep fields \citepalias{DES_DF_Y3}. Estimates of stellar contamination from different subsets of a catalog can vary slightly from each other; for example, our estimates using \Balrog are slightly different from the estimates from the Y6 Gold catalog \citep[][see their Figure 3]{Y6Gold} which matches DES objects to a combination of data from the Vista Hemisphere Survey \citep[VHS DR5,][]{McMahon:2013} and HSC \citep{Miyazaki:2018:HSC}.

Figure \ref{fig:StarGalSep} shows our results. The four leftmost panels present the galaxy efficiency and stellar contamination for the different galaxy samples. The contamination in the two source samples is below $\lesssim 1\%$ (shown as the horizontal gray band) across the full magnitude range $18 < m_i < 25$. For the lens samples, the contamination is below $1\%$ for the brighter magnitudes of the \maglim sample, and similarly for the \redmagic high-density sample. The contamination also rises towards fainter magnitudes. This is expected as the number of selected objects in a given sample (the denominator in the computed fractions) decreases significantly at fainter magnitudes; this is especially true for the lens samples, which have only a few fainter objects. The decreasing denominator results in a larger stellar fraction. The background of each panel presents a histogram of the number of objects at a given $i$-band magnitude. As expected, any increases in stellar contamination fractions are in magnitude ranges that are at the faint tail of the distribution for each of the four main galaxy samples. 

The source samples are the most efficient in their galaxy selection, as they discard the fewest true galaxies in their selection. The lens samples' selections are far more restrictive, with \redmagic being the most restrictive. This is expected as the lens samples select specific types of galaxies that are best suited for obtaining precise redshift estimates. We do not present an efficiency estimate for the \mdet sample as we do not store the necessary data (i.e. the full catalog, without any selections applied) due to storage limitations; recall that \mdet provides five different detection catalogs, which increases our storage footprint by five times compared to other samples.

Finally, the rightmost panels of Figure \ref{fig:StarGalSep} show the same estimates but for the full Y6 Gold sample, from varying the \fitvd-based star-galaxy classifier \citep{Y6Gold}. Note that this classifier is used in defining the lens samples, but not the source samples. The panel shows that $\texttt{EXT\_MASH} \geq 4$ results in a percent-level contamination rate. The bottom right panel shows an inverted estimate, where we compute the \textit{galaxy} contamination and the \textit{stellar} efficiency. Such estimates are valuable for any science derived from the DES star catalogs. Our results affirm the findings in previous DES works \citep[\eg][]{BalrogY3, PhotoSet_Cosmo_Y3}, that the morphological classifier still results in significant ($> 10\%$) contamination from galaxies when selecting a star sample. In DES Y6, we also develop an improved morphology-based estimator --- based on the gradient-boosted decision tree algorithm, \textsc{XGBoost} \citep{Chen:2016} --- which enhances stellar science by selecting purer star samples from the Y6 data \citep[][see their Section 4.2 and Appendix A.2]{Y6Gold}. This classifier is not implemented for the \Balrog catalog as it requires an additional postprocessing step that is not part of our main analysis.

We make a brief note about two subtleties of the analysis in this section. First, the accuracy of star-galaxy classification can suffer if there are non-negligible errors in the PSF model. For example, if the true PSF is larger/smaller than the PSF model, then classifiers that use the size of the object and/or the size of the PSF would incur errors. This error will not be observed in \Balrog, which uses the same PSF for convolving a source during injection and for deconvolving observations of the source during model fitting. Note that \citet{Y6PSF} (see their tests in Section 4) show that the PSF model in Y6 reproduces the size of the true PSF (estimated using observed stars) within 1\%, and the ellipticities within $\delta e \sim 5 \times 10^{-4}$. Thus, this subtlety is not a notable concern.

Second, stars in partially resolved binary systems will be measured as slightly extended objects, and can therefore be incorrectly classified as galaxies during the star-galaxy classification. \Balrog will include such binaries as such sources will be present in the deep-field catalog from which we draw our injections. However, their impact of the stellar contamination crucially relies on accurate classification of such binary systems in the deep-field catalog (given this classification is used to determine the stellar contamination). We do not estimate the accuracy of the color-based estimator from \citetalias{DES_DF_Y3} in classifying partially resolved binaries.

\section{Applications to DES Y6 Projects}\label{sec:applications}

The \Balrog pipeline and its derived data products are necessary for key pieces of the DES cosmology analysis (\eg redshift calibration and magnification estimates), and for informing the quality of our photometric/pipeline datasets. We have thus far detailed the latter in the section above, and now discuss a few examples of the former, with specific use-cases in DES Y6.

\subsection{Photometric redshift calibration}\label{sec:sec:SOMPZ}

A key application of \Balrog is in the photometric redshift calibration of the source sample and the lens sample \citep{Myles:2021:SOMPZ, Sanchez:2023:highz, Giannini:2024:DES}. The main DES $3\times2$ point analysis --- which uses galaxy clustering, cosmic shear, and galaxy-galaxy lensing to infer cosmology --- relies on a robust characterization the source and lens samples' redshift distributions in order to infer cosmology \citep{DES_y1_Cosmology, DES:2022:3x2pt_Y3}. The characterizations for Y6 will be presented in \citet{Y6SOMPZmdet, Y6SOMPZBFD, Y6SOMPZMaglim} for the \mdet, BFD, and \maglim sample, respectively. A significant challenge in estimating redshifts in photometric surveys is the presence of degeneracies in the color-redshift relation. These degeneracies can be broken by including additional photometric measurements per galaxy, but this is not always available for all galaxies in the full DES footprint. Following \citet{Buchs_2019}, DES has used a fully Bayesian technique for learning the color-redshift relation in regions of the sky with low-noise, multi-band photometry measurements, and then transferring that knowledge to all galaxies across the full DES footprint.

In our case, the low-noise data is obtained from the DES Y3 deep fields \citepalias{DES_DF_Y3}, which are four patches of the sky totaling $\approx 5.9 \deg^2$ that are observed to significantly greater depth (i.e., the photometric measurements have lower noise). A majority of galaxies in these patches also have measurements in eight bands ($u, g, r, i, z, J, H, K_s$), as opposed to the four bands of the wide field ($g, r, i, z$), due to supplemental data from other surveys \citep[\eg][]{VIDEO, VVDS_2013}. A large fraction of these available galaxies also have redshift estimates, provided by other surveys. The color-redshift relation is learned in these deep fields, and then extrapolated to the fiducial galaxy samples, which only have measurements in four bands. \Balrog provides a robust, probabilistic way of performing this conversion, while accounting for the complexities in the images. Furthermore, it provides a natural way to calibrate the biases between recovered photometry and true photometry, i.e. the offsets discussed in Section \ref{sec:sec:performancefid} are naturally incorporated in our redshift estimation through the use of \Balrog.

In brief, the probability distribution for the redshift of a given set of galaxies can be written as
\begin{equation}\label{eqn:SOMPZ}
    p(z|\hat{c},\hat{s})=
    \sum_c
    p(z|c,\hat{c},\hat{s})p(c|\hat{c},\hat{s})\,,
\end{equation}
where $z$ is redshift, $c$ is a galaxy phenotype, defined using the deep-field measurements (the 8-band, low-noise photometry); $\hat{c}$ is the phenotype defined using the wide-field measurements (the 4-band, noisy photometry), and $\hat{s}$ is the sample selection function. The second factor, $p(c|\hat{c},\hat{s})$, is the transfer function, which is measured with \Balrog. This quantity represents the probability of a galaxy being in deep-field phenotype $c$, given it is already in wide-field phenotype $\hat{c}$ and has passed the selection $\hat{s}$. We use ``phenotype'' to refer to the classification of a galaxy into a low-dimensional space using data from a higher-dimensional space. As a simple example, galaxies can be classified into red and blue ``phenotypes'' based on their photometric measurements. The classification step in DES Y6 is performed using self-organizing maps (SOMs), following \citet{Buchs_2019, Myles:2021:SOMPZ}, which is an unsupervised, data-driven algorithm.

In practice, we estimate redshift distributions for galaxies placed in different tomographic bins. In this case, the redshift distribution in each tomographic bin $\hat{b}$ is estimated as
\begin{equation}
    p(z|\hat{b},\hat{s})=
    \sum_{\hat{c} \in \hat{b}} \sum_c p(z|c,\hat{b},\hat{s})p(c|\hat{c},\hat{b},\hat{s})p(\hat{c} | \hat{b}, \hat{s})\,,
\end{equation}
The \Balrog data is used for assigning galaxies of phenotype $\hat{c}$ to tomographic bins, by computing the mean redshift of all \Balrog galaxies in a given phenotype \citep[\eg][]{Myles:2021:SOMPZ, Sanchez:2023:highz, Giannini:2024:DES}. Then, \Balrog provides a transfer function --- which is the probability of a galaxy of wide-field phenotype $\hat{c}$ also being in deep-field phenotype $c$ --- corresponding to all galaxies in a given tomographic bin.

We have already presented a rough version of the transfer function in Section \ref{sec:sec:performancefid}. The phenotypes $c$ and $\hat{c}$ are discretizations of the deep-field and wide-field photometry, respectively. Figures \ref{fig:y6deepmags} and \ref{fig:y6deepcolor} show the distributions of measured fluxes/colors as a function of the true, noiseless quantities. This can be thought of as one slice of the distribution $p(\hat{c} | c)$. Those figures show that for fainter objects (which dominate the sample, given their larger number density relative to that of bright objects) there can be large differences between the true, deep-field photometry and the observed, wide-field photometry. Such behaviors are accurately captured by computing the transfer function through \Balrog.

\citet[][see their Section 5.4]{Myles:2021:SOMPZ} have also already shown that the DES Y3 \Balrog dataset contributes an infinitesimal uncertainty to the final redshift distribution estimates. We expect this contribution to be even smaller in Y6, as the previous estimate accounted for the fact that the Y3 \Balrog tiles only sampled $\approx 20\%$ of the DES Y3 footprint. In Y6, we sample the entire footprint and therefore remove this uncertainty factor.

\subsection{Lensing magnification}\label{sec:sec:mag}

\begin{figure}
\includegraphics[width = \columnwidth]{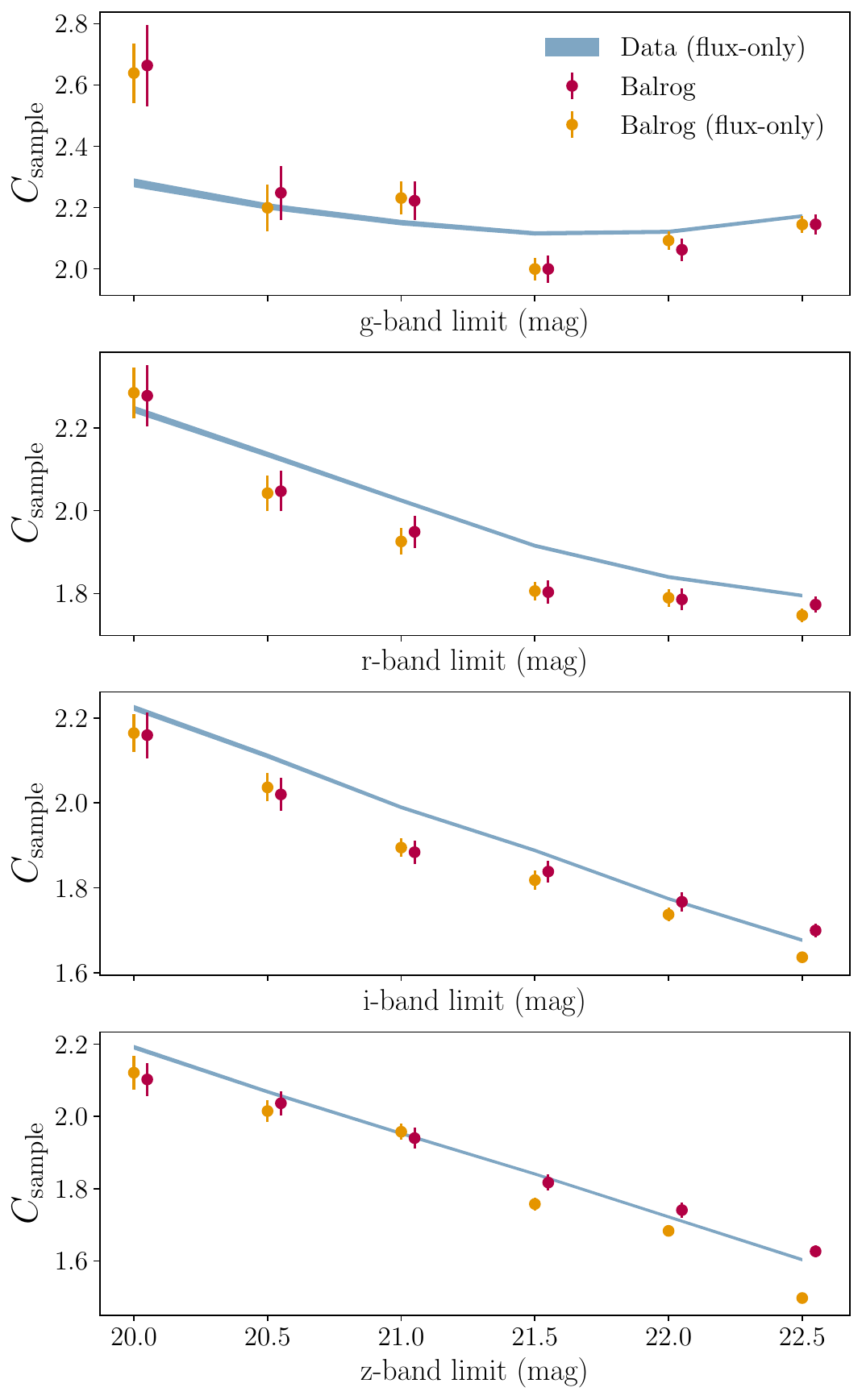}
\caption{Estimates of the magnification bias $C_{\mathrm{sample}}$ from boosted detection efficiency in samples defined with a flux magnitude limit (shown in the x-axis). The \Balrog estimates (red) use the magnified runs, in which a 2\% magnification is applied to every input object and the entire image processing is rerun. The data estimate (blue) applies the same magnification to galaxy magnitudes from the Y6 Gold sample and re-applies the same selection function. Finally, we show a \Balrog-based flux-only estimate for comparison. In all cases, the error for each magnitude-limited sample is obtained via a jackknife estimate over 100 patches of the footprint.}
\label{fig:magnification}
\end{figure}

Another key application of \Balrog is in the estimation of magnification coefficients. Gravitational lensing from the foreground matter distribution induces a magnification on the background structure \citep{Bartelmann_2001}. This alters the observed surface area from a given redshift, and also impacts the measured properties of galaxies at that redshift, such as flux and area, which in turn changes the number of galaxies that pass a given selection function. As a result, magnification impacts any analyses that relies of the counts of galaxies, such as the galaxy clustering and galaxy-galaxy lensing analyses, as shown in DES Y3 \citep{Magnification_Y3}. It is also a key input in one of the redshift calibration methods, using the clustering of galaxies \citep{GattiGiannini:2022:Wz, Y6WZ}.

Magnification can be expressed in terms of the convergence $\kappa$ and the shear $\gamma$ \citep{Bartelmann_2001}. In the weak lensing regime, $\kappa \ll 1$ and $\gamma \ll 1$ --- where $\kappa$ is the lensing convergence, and $\gamma$ is the two-component shear --- so the local magnification can be written as
\begin{equation}
    \mu=\frac{1}{(1-\kappa)^2-|\gamma|^2}
    \approx \frac{1}{1-2\kappa}
    \approx 1+2\kappa\,.
\end{equation}
Given such a magnification, we can rewrite the observed galaxy number density field $\delta_g^{\mathrm{obs}}$ into two additive components: an intrinsic density fluctuation, and a contribution due to magnification,
\begin{equation}
    \delta_g^{\mathrm{obs}} = 
    \delta_g^{\mathrm{int}} + \delta_g^{\mathrm{mag}}\,.
\end{equation}
As mentioned before, the contribution from magnification is the result of two separate and competing effects: an increase in the observed area element that decreases the observed number density of galaxies, and an enhancement in the detection efficiency of faint galaxies that increases the observed number density. With these effects, the observed local density fluctuation is given by
\begin{equation}
    \delta_g^{\mathrm{mag}} = 
    \left(C_{\mathrm{sample}} + C_{\mathrm{area}}\right)\delta\kappa\,,
\end{equation}
where $C_{\mathrm{sample}}$ is the factor associated to the boost in detection efficiency and $C_{\mathrm{area}}$ is the factor related to the decrease in number density. We can set $C_{\mathrm{area}} \equiv -2$ given the terms encapsulates a known geometric effect \citep{Magnification_Y3} whereas $C_{\mathrm{sample}}$ depends on the exact definition of the sample selection function. To linear order, it can be estimated as the numerical derivative
\begin{equation}\label{csample1}
    C_{\mathrm{sample}}\delta\kappa=
    \frac{N_{\delta\kappa}-N_0}{N_0}\,,
\end{equation}

\noindent where $N_{\delta\kappa}$ is the number of objects, whose properties have been impacted by magnification $\delta \kappa$, that pass the sample selection function. Then $N_0$ is the number of objects, without any magnification, that pass the selection function.

To estimate $C_{\mathrm{sample}}$, we run a magnified variant of \Balrog for 2000 tiles $(1,000 \deg^2)$, or $20\%$ of the full DES Y6 footprint, where each variant tile has the exact same injections (including the same positions on the sky) as in the fiducial version but now with a $2\%$ increase in the flux and area of all injected objects, corresponding to magnification due to a lensing convergence of $\delta\kappa=0.01$. Given a galaxy sample with a specific selection function, we can calculate $C_{\mathrm{sample}}$ from Equation \eqref{csample1}, with $N_{\delta\kappa}$ measured on the magnified runs and $N_0$ measured on the fiducial runs. An approximate estimate of this method can also be made using a ``flux-only'' method where one takes the fiducial catalog, then increases the measured flux by $2\%$ and applies the selection function of interest. This method will only trace a subset of the selection effects captured by the \Balrog-based estimate.

In Figure \ref{fig:magnification} we show estimates of $C_{\rm sample}$, where the selection function is just a simple magnitude threshold in a given photometric band. We make three different estimates, using two samples and two techniques. In all cases, we only use galaxies that pass the Y6 Gold selection and the star-galaxy selection; our selection function therefore includes a magnitude limit \textit{and} a star-galaxy selection. We estimate $C_{\rm sample}$ using the fiducial and magnified \Balrog samples, which contain the full impact of magnification on all measured properties of a galaxy. We then also perform a flux-only magnification using the fiducial \Balrog catalog and then also using the data catalog. In all cases, the error is obtained by performing a jackknife estimate over 100 patches of the footprint.

We present all three estimates in Figure \ref{fig:magnification}. The \Balrog magnification estimates are consistent with the \Balrog flux-only estimates, with some discrepancies at the faint end in redder bands. These differences reflect the limitation in using a flux-only method to estimate the full impact of magnification on the sample selection function. Next, the data flux-only estimate is generally similar to, but statistically distinct from, the \Balrog flux-only estimate. While the \Balrog data is largely representative of the real data, differences in the distribution of object properties between the two samples will cause differences in the estimates of $C_{\mathrm{sample}}$. Figure \ref{fig:magnification} shows that such differences are at the few percent level for most of the datapoints. The main magnification estimates in DES Y6 find that such \Balrog--data differences are subdominant to the uncertainties of the $C_{\mathrm{sample}}$ estimates \citep{Y6Mag}.

\subsection{Depth estimates}\label{sec:sec:detectioneff}

In Figure \ref{fig:DetectionRate} (and Section \ref{sec:sec:sec:completeness}) we show the detection efficiency for galaxies of different brightness, in different bands. We have motivated above how this efficiency, especially as a function of the object's true properties, is relevant for many science cases in DES Y6; with the redshift estimation (Section \ref{sec:sec:SOMPZ}) and magnification estimates (Section \ref{sec:sec:mag}) being two pertinent examples. It is also useful to quantify the raw completeness of the general purpose catalog. In DES Y3 this was done through a variety of techniques, one of which was to cross-match objects from the DES deep fields with the same object in the wide-field data (which is shallower) and compute a detection efficiency accordingly. This was accompanied by a \Balrog-derived estimate as well. The latter is a purer test as we have exact control on the objects being injected into the data, but retain the realism of the DES dataset by injecting into real images. Similarly, in DES Y6, the \Balrog dataset presented here is used to characterize the exact detection efficiency of the survey data; particularly the magnitude limit, per band, at which a galaxy is detected $90\%$ of the time.

In Table \ref{tab:completeness} we present an estimate of survey depth, using the Y6 \Balrog and Y6 Gold samples, for sources with $S/N = 10$. This is computed through the following process: (1) selecting all objects with $9.5 < \texttt{flux} / \texttt{flux\_err} < 10.5$ in a given band, and with $\texttt{EXT\_MASH} = 4$ so that we select only  extended objects (2) computing the mean magnitude in pixels of a map with $\texttt{NSIDE} = 256$, and (3) computing the median and 16/84\% percentile values of this magnitude over the footprint. The magnitude limits for the two samples are consistent at the 2-3\% level, which is significantly smaller than the variation of this limit across the sky.

\begin{table}
    \centering
    \begin{tabular}{c|c|c}
        band & \Balrog & Y6 \textsc{Gold} \\
        \hline
        $g$ & $24.14^{+0.21}_{-0.19}$ & $24.16^{+0.11}_{-0.09}$ \\
        $r$ & $23.82^{+0.20}_{-0.17}$ & $23.84^{+0.11}_{-0.09}$ \\
        $i$ & $23.26^{+0.20}_{-0.18}$ & $23.28^{+0.11}_{-0.09}$\\
        $z$ & $22.61^{+0.21}_{-0.20}$ & $22.62^{+0.12}_{-0.10}$\\
        \hline
    \end{tabular}
    \caption{The magnitude limit for extended objects with $9.5 < \texttt{flux} / \texttt{flux\_err} < 10.5$ in a given band, computed using \fitvd quantities from the \Balrog synthetic data and the Y6 Gold data. The limits estimated from the two samples are consistent at the 2-3\% level. The \Balrog estimate is noisier as the sample contains fewer objects relative to Y6 Gold. See text for details on the calculation.}
    \label{tab:completeness}
\end{table}

\subsection{Large-scale survey systematics}\label{sec:sec:LSSweights}

Observations of the galaxy number density field, and in particular of the spatial correlations in this field, are sensitive to many cosmological processes \citep{Peebles:1980} and have been used to extract constraints on parameters \citep[\eg][]{GilMarin:2015:BOSS, Alam:2017:BOSS, Alam:2021:eBOSS, DES:2022:3x2pt_Y3}. As a result, they have been extensively used in the DES Y1 and Y3 analyses \citep{DES_y1_Cosmology, DES:2022:3x2pt_Y3}. A critical part of such analyses is the characterization, and subsequent correction, of spatial correlations in the field that arise from non-cosmological sources, such as the observing conditions of the survey. For example, we have shown in Figure \ref{fig:footprint} that the number of detected galaxies (either in \Balrog or the Y6 Gold data) is lower near the east and west edges of the DES footprint, where the stellar density is higher. This causes a spatial correlation in the number density that does not originate from any cosmological process.

Such non-cosmological correlations can be corrected by assigning weights to the galaxies, where the weights are determined using the survey property maps (see Section \ref{sec:sec:sec:sysmaps}) in conjunction with a variety of data-driven methods \citep[\eg][]{Y3_GalClustering}. \Balrog provides a powerful test dataset for the robustness of these methods. The synthetic sources in \Balrog are injected on a uniform grid (see Section \ref{sec:InjWgtscheme}). Under ideal conditions with no systematic effects, the number density of synthetic sources will have no correlation with that of the real sources in the data. However, observational effects will necessarily generate correlations between the two samples, as both samples are impacted by the presence of such observational effects in the real images. Vice-versa, if these non-cosmological correlations are completely corrected for, then there will be no spatial correlations between the two samples. Thus, \Balrog can provide a vital estimate of the residual systematics in the galaxy correlation measurements. Such a test was performed in DES Y3 and was one aspect of the lens sample validations \citep[][see their Section 8.5 and Figure 12]{Y3_GalClustering}. The same validation in Y6 will be significantly more precise given the factor of 5 increase in area covered by \Balrog and a factor of 12 increase in number of usable synthetic lens galaxies.

\section{Summary and Conclusion}\label{sec:conclusion}

Photometric surveys are powerful tools for studying astrophysics and cosmology across many epochs and scales. In this work, we present the synthetic source catalogs used to characterize the DES Y6 image processing pipelines and the resulting galaxy samples derived from the catalogs. The synthetic sources are injected into real CCD images from DES Y6, and the entire processing pipeline is rerun on these source-injected images. This work is a supporting analysis for the main DES Y6 cosmology efforts. The \Balrog catalog from DES Y6 is also currently the largest synthetic catalog at this fidelity level, and enables new science cases. We list our main results/findings below:

\begin{itemize}
    \item \Balrog in DES Y6 now provides synthetic sources for the entire \(\sim\)$5000 \deg^2$ footprint of the survey, which both increases the number of injections over Y3 (from $\approx 25$ million to $146$ million) and also enables more precise analyses of area-dependent effects, as we point out below.
    
    \item A ``tiered'' injection scheme for the synthetic sources greatly increases the size of the \Balrog sample relevant for the cosmology analysis, while still injecting a large/representative sample of sources. For source and lens samples, we find increases of $300\%$ and $1200\%$, respectively (Figure \ref{fig:weightscheme}).

    \item The synthetic sample is representative of the data (Figure \ref{fig:CompareMdet}) and accurately captures the correlation between galaxy properties and observing conditions of the survey (Figure \ref{fig:survey_properties} and \ref{fig:survey_properties_maglim}). The latter is not constructed by hand, and instead manifests directly from the complexities of the real images that \Balrog injects synthetic sources into.

    \item Through \Balrog, we show the DES Y6 image processing pipeline can recover the mean photometry (averaged over many detections) at the percent level (Figure \ref{fig:y6deepmags} and \ref{fig:y6deepcolor}), with deviations for the faintest objects and reddest bands that is consistent with an (expected) detection bias and potential background subtraction biases from \SE.

    \item The photometric accuracy is similar for stars, which have a much simpler morphology than galaxies, with a slightly better accuracy in general (Figure \ref{fig:starmags} and \ref{fig:starcolor}).

    \item The star-galaxy classification of the general purpose \fitvd catalog, and of the two source samples and two lens samples, is found to limit stellar contamination to the percent level or lower over the magnitude range of interest (Figure \ref{fig:StarGalSep}).
\end{itemize}

This work has focused on the use-cases of \Balrog that are particularly relevant for the DES Y6 cosmology analysis. There are also other use-cases beyond those discussed above: for example, \Balrog has been used to study the impact of unresolved sources and background subtraction \citep{Eckert:2020}, to learn the distribution of observed properties as a function of survey conditions, to improve precision measurements of host galaxy substructure \citep{Doliva-Dolinsky:2025:NGC3109}, etc. 

SSI has become a critical tool to robustly characterize and use data from surveys. In the upcoming decade, the Vera C. Rubin Observatory's Legacy Survey of Space and Time (LSST) will observe over 20,000 $\deg^2$ of the sky, and will achieve higher precision in photometric and morphological measurements. SSI pipelines, similar to the \Balrog pipeline used in this work, will therefore be necessary for most survey tasks. The dataset presented in this work is the largest synthetic source catalog built for a photometric survey and therefore will prove a viable testing ground for how SSI datasets can benefit/complement analyses with LSST data.

\section*{Author Contributions}
DA, MT and BY worked on the pipeline for \Balrog DES Y6 --- which was based on the original pipeline of DES Y3 from SE --- with assistance from MB, MY, GB, RG, MJ and ES for various technical details/improvements of the processing pipeline done in Y6. DA and MT generated the entire \Balrog dataset and produced all the resulting catalogs. DA and JB performed all analyses, and wrote all the text, as presented in this work. KB and JE served as our internal review committee and contributed to the writing and presentation of this paper. The \maglim sample for \Balrog was generated by EL and GG, and the DNF redshift catalog by JD. The Y6 \mdet pipeline run on \Balrog (as well as data) was developed by MY, MB, ES, MJ, RG, ERy and TS. The same for BFD was developed by MG, GB, VW, and MJ. The \maglim sample definitions was constructed by NW, MR, SL, JE, AC, JM and AP. The survey property maps we use were developed by ERy, while the \redmagic sample selection was defined by ERy and ERo. The remaining authors have made contributions to this paper that include, but are not limited to, the construction of DECam and other aspects of collecting the data; data processing and calibration; developing broadly used methods, codes, and simulations; running the pipelines and validation tests; and promoting the science analysis.

\section*{Acknowledgements}

DA is supported by the National Science Foundation Graduate Research Fellowship under Grant No.\, DGE 1746045. DA also thanks Richard Kron for discussions on the history of photometric surveys and observational techniques.

Funding for the DES Projects has been provided by the U.S. Department of Energy, the U.S. National Science Foundation, the Ministry of Science and Education of Spain, 
the Science and Technology Facilities Council of the United Kingdom, the Higher Education Funding Council for England, the National Center for Supercomputing 
Applications at the University of Illinois at Urbana-Champaign, the Kavli Institute of Cosmological Physics at the University of Chicago, 
the Center for Cosmology and Astro-Particle Physics at the Ohio State University,
the Mitchell Institute for Fundamental Physics and Astronomy at Texas A\&M University, Financiadora de Estudos e Projetos, 
Funda{\c c}{\~a}o Carlos Chagas Filho de Amparo {\`a} Pesquisa do Estado do Rio de Janeiro, Conselho Nacional de Desenvolvimento Cient{\'i}fico e Tecnol{\'o}gico and 
the Minist{\'e}rio da Ci{\^e}ncia, Tecnologia e Inova{\c c}{\~a}o, the Deutsche Forschungsgemeinschaft and the Collaborating Institutions in the Dark Energy Survey. 

The Collaborating Institutions are Argonne National Laboratory, the University of California at Santa Cruz, the University of Cambridge, Centro de Investigaciones Energ{\'e}ticas, 
Medioambientales y Tecnol{\'o}gicas-Madrid, the University of Chicago, University College London, the DES-Brazil Consortium, the University of Edinburgh, 
the Eidgen{\"o}ssische Technische Hochschule (ETH) Z{\"u}rich, 
Fermi National Accelerator Laboratory, the University of Illinois at Urbana-Champaign, the Institut de Ci{\`e}ncies de l'Espai (IEEC/CSIC), 
the Institut de F{\'i}sica d'Altes Energies, Lawrence Berkeley National Laboratory, the Ludwig-Maximilians Universit{\"a}t M{\"u}nchen and the associated Excellence Cluster Universe, 
the University of Michigan, NSF NOIRLab, the University of Nottingham, The Ohio State University, the University of Pennsylvania, the University of Portsmouth, 
SLAC National Accelerator Laboratory, Stanford University, the University of Sussex, Texas A\&M University, and the OzDES Membership Consortium.

Based in part on observations at NSF Cerro Tololo Inter-American Observatory at NSF NOIRLab (NOIRLab Prop. ID 2012B-0001; PI: J. Frieman), which is managed by the Association of Universities for Research in Astronomy (AURA) under a cooperative agreement with the National Science Foundation.

The DES data management system is supported by the National Science Foundation under Grant Numbers AST-1138766 and AST-1536171.
The DES participants from Spanish institutions are partially supported by MICINN under grants PID2021-123012, PID2021-128989 PID2022-141079, SEV-2016-0588, CEX2020-001058-M and CEX2020-001007-S, some of which include ERDF funds from the European Union. IFAE is partially funded by the CERCA program of the Generalitat de Catalunya.

We  acknowledge support from the Brazilian Instituto Nacional de Ci\^encia
e Tecnologia (INCT) do e-Universo (CNPq grant 465376/2014-2).

This manuscript has been authored by Fermi Research Alliance, LLC under Contract No. DE-AC02-07CH11359 with the U.S. Department of Energy, Office of Science, Office of High Energy Physics.

\bibliography{Paper}
\bibliographystyle{aasjournal}

\appendix

\section{Correlations of lens galaxies with survey properties: \maglim \& \redmagic}\label{appx:SPmaps_Lens}

\begin{table*}
    \centering
    \begin{tabular}{|c|c|c|c|c|c|c|c|c|}
    \hline
    Magnitude & \multicolumn{4}{c|}{Galaxies (mmag)} & \multicolumn{4}{c|}{Stars (mmag)}\\
    \hline
    & $g$ & $r$ & $i$ & $z$ & $g$ & $r$ & $i$ & $z$ \\
    \hline
    17.0 - 17.25 & -8 (31) & -4 (22) & -4 (18) & -3 (18) & --- & --- & --- & --- \\
    17.25 - 17.5 & -12 (31) & -12 (23) & -7 (19) & -2 (18) & --- & --- & --- & --- \\
    17.5 - 17.75 & -27 (31) & -9 (23) & -2 (19) & -2 (20) & --- & --- & --- & --- \\
    17.75 - 18.0 & -8 (30) & -9 (23) & -5 (21) & -1 (22) & --- & --- & --- & --- \\
    18.0 - 18.25 & -6 (30) & -6 (24) & -4 (22) & -1 (23) & -23 (21) & -9 (19) & -3 (14) & 0.5 (12)\\
    18.25 - 18.5 & -12 (32) & -6 (25) & -3 (23) & -0.6 (24) & -10 (27) & -8 (19) & -2 (15) & 1 (13)\\
    18.5 - 18.75 & -9 (30) & -7 (26) & -3 (25) & 0.7 (27) & -23 (28) & -0.1 (19) & -2 (15) & 3 (13)\\
    18.75 - 19.0 & -12 (32) & -8 (27) & -2 (26) & 0.1 (28) & -0.9 (28) & -1 (18) & 1 (15) & 1 (14)\\
    19.0 - 19.25 & -7 (32) & -6 (28) & -2 (28) & 0.9 (31) & -7 (28) & -7 (21) & -0.2 (16) & 3 (14)\\
    19.25 - 19.5 & -12 (33) & -7 (29) & -3 (30) & 2 (34) & -3 (27) & -4 (19) & -0.1 (16) & 3 (15)\\
    19.5 - 19.75 & -11 (34) & -6 (31) & -1 (32) & 3 (37) & -20 (29) & -6 (21) & -0.7 (17) & 3 (17)\\
    19.75 - 20.0 & -12 (35) & -7 (33) & -0.7 (36) & 4 (41) & -14 (28) & -4 (20) & -0.2 (17) & 3 (18)\\
    20.0 - 20.25 & -13 (37) & -6 (34) & 0.1 (39) & 6 (47) & -18 (30) & -4 (22) & -0.1 (18) & 4 (19)\\
    20.25 - 20.5 & -11 (38) & -6 (37) & 0.6 (43) & 7 (52) & -17 (30) & -3 (21) & 0.2 (19) & 5 (21)\\
    20.5 - 20.75 & -14 (40) & -5 (41) & 1 (48) & 8 (59) & -17 (30) & -4 (23) & 1 (21) & 6 (24)\\
    20.75 - 21.0 & -13 (43) & -5 (45) & 2 (53) & 9 (66) & -20 (31) & -5 (24) & 0.2 (22) & 7 (27)\\
    21.0 - 21.25 & -14 (45) & -4 (50) & 3 (59) & 9 (73) & -18 (33) & -5 (25) & 1 (24) & 9 (31)\\
    21.25 - 21.5 & -14 (49) & -4 (55) & 2 (66) & 11 (84) & -16 (33) & -5 (26) & 4 (26) & 10 (36)\\
    21.5 - 21.75 & -14 (53) & -4 (61) & 3 (74) & 13 (95) & -13 (34) & -4 (28) & 5 (30) & 13 (44)\\
    21.75 - 22.0 & -14 (59) & -4 (69) & 5 (83) & 17 (110) & -16 (36) & -4 (31) & 6 (35) & 16 (52)\\
    22.0 - 22.25 & -15 (66) & -2 (77) & 8 (94) & 23 (127) & -16 (37) & -0.8 (34) & 11 (39) & 21 (63)\\
    22.25 - 22.5 & -15 (73) & -0.8 (87) & 13 (107) & 32 (148) & -15 (40) & -0.6 (39) & 14 (47) & 27 (76)\\
    22.5 - 22.75 & -15 (83) & 2 (98) & 20 (123) & 42 (175) & -12 (45) & 3 (44) & 19 (56) & 31 (95)\\
    22.75 - 23.0 & -14 (94) & 8 (113) & 29 (143) & 51 (207) & -13 (50) & 6 (51) & 25 (67) & 38 (114)\\
    23.0 - 23.25 & -12 (108) & 15 (130) & 39 (168) & 58 (246) & -10 (57) & 10 (62) & 30 (83) & 47 (140)\\
    23.25 - 23.5 & -7 (124) & 24 (150) & 47 (199) & 56 (291) & -8 (68) & 15 (73) & 40 (102) & 55 (177)\\
    23.5 - 23.75 & -1 (144) & 33 (176) & 48 (237) & 33 (340) & -6 (82) & 22 (90) & 50 (130) & 59 (223)\\
    23.75 - 24.0 & 6 (168) & 38 (207) & 30 (278) & -19 (386) & -4 (98) & 29 (110) & 63 (158) & 58 (271)\\
    24.0 - 24.25 & 11 (198) & 28 (245) & -21 (322) & -113 (423) & -1 (119) & 42 (135) & 74 (197) & 31 (316)\\
    24.25 - 24.5 & 6 (233) & -12 (291) & -117 (360) & -250 (440) & 6 (145) & 51 (169) & 77 (241) & -36 (346)\\
    24.5 - 24.75 & -13 (279) & -93 (339) & -252 (384) & -404 (429) & 13 (180) & 67 (205) & 67 (285) & -176 (367)\\
    24.75 - 25.0 & -51 (333) & -210 (380) & -402 (387) & -543 (395) & 25 (221) & 82 (253) & 3 (329) & -331 (371)\\
    \hline
    \end{tabular}
    \vspace{10pt}
    \caption{The median difference between the true and measured magnitudes as shown in Figure \ref{fig:y6deepmags} and \ref{fig:starmags}, in units of millimags. The numbers in the parantheses show the standard deviation of the differences, shown as black bars in the aforementioned figures. Our synthetic star sample contains too few objects below $m < 18$ so we do not tabulate the bias there.}
    \label{tab:photo_acc}
\end{table*}

\begin{figure*}[!h]
\centering
\includegraphics[width = 1.8\columnwidth]{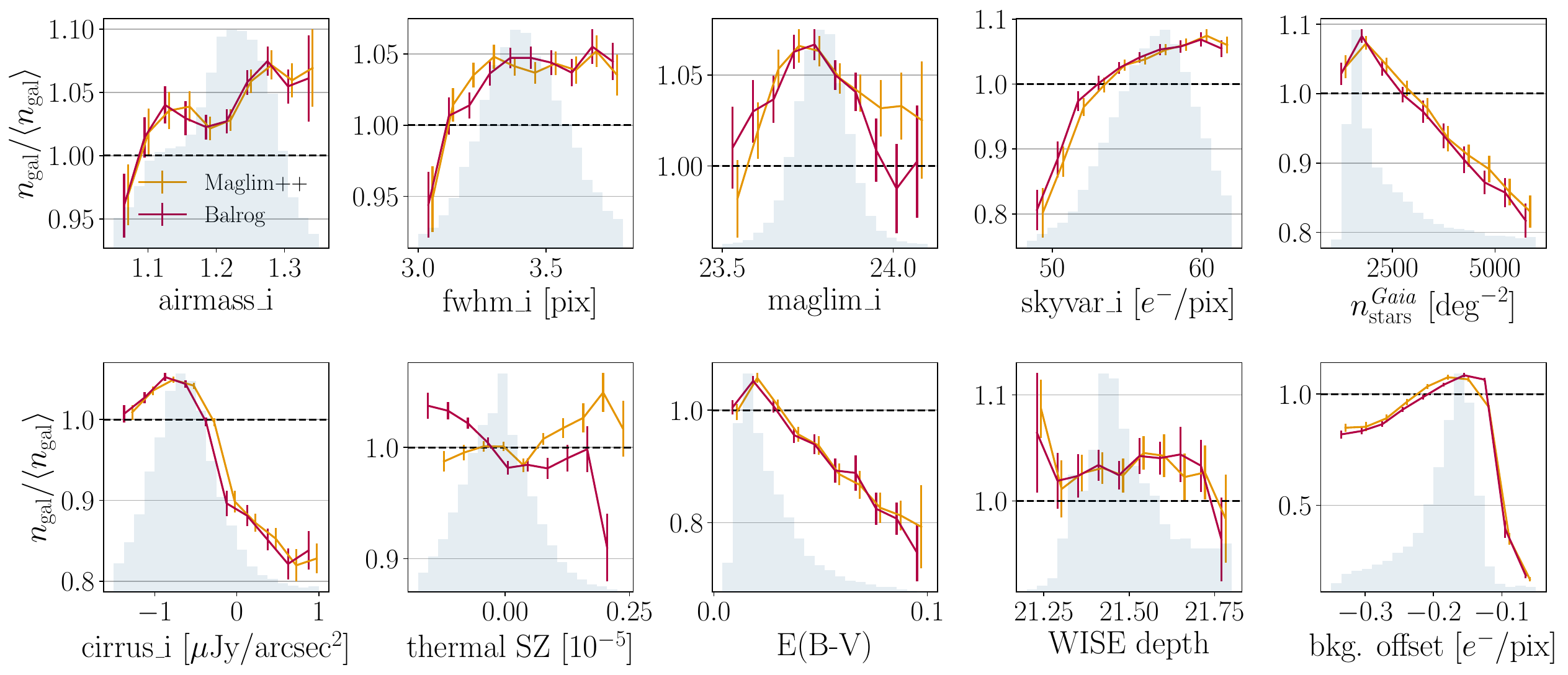}
\par\vspace{0.7cm} 
\includegraphics[width = 1.8\columnwidth]{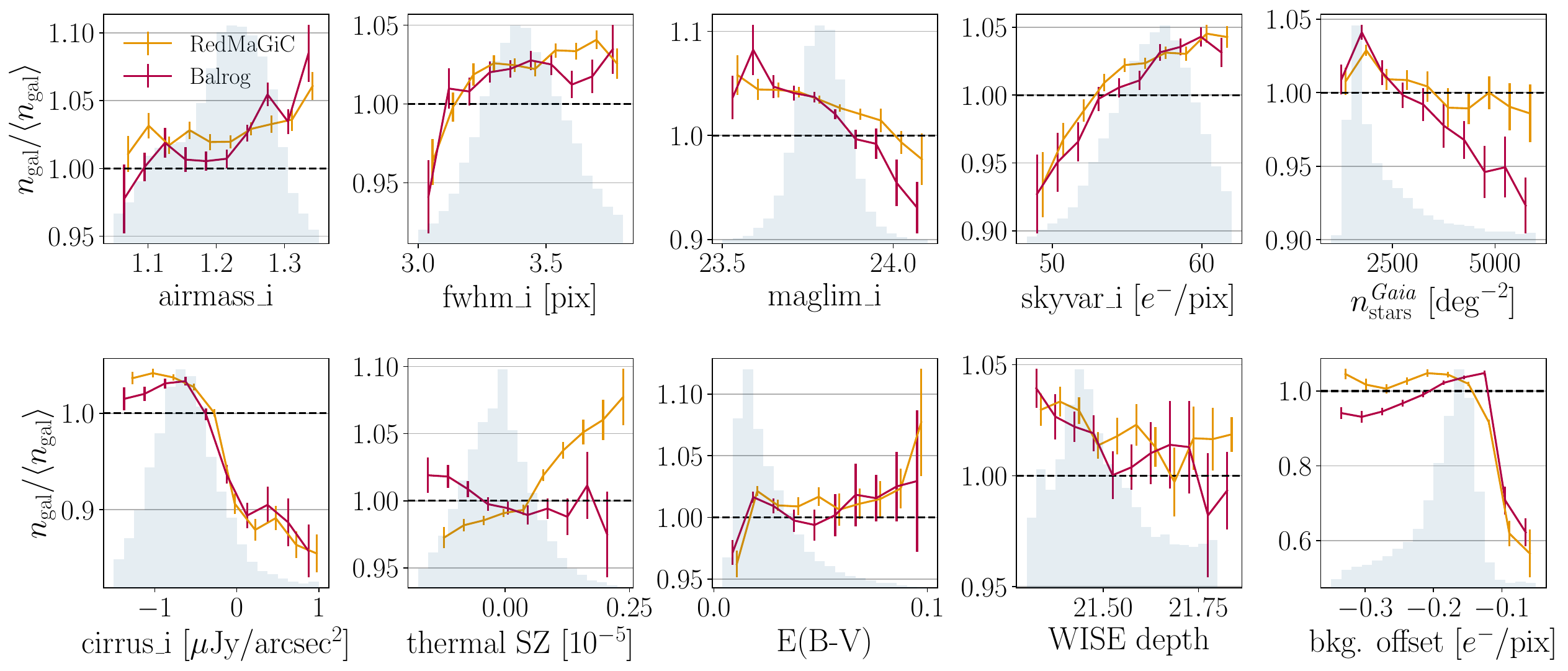}
\caption{The trend in number density fluctuations $N/\langle N\rangle$ as a function of various survey observing properties for the \Balrog (red) and data (orange) samples. Analogous to Figure \ref{fig:survey_properties} but now for the \maglim (top) and \redmagic (bottom) sample. We use $\texttt{NSIDE = 128}$ as this is a smaller sample than the Gold catalog used in Figure \ref{fig:survey_properties} and so covers the footprint in a more sparse manner. The $\chi^2$ between the measurements on data and \Balrog (for either lens sample) exhibits $\chi^2 / N_{\rm dof} \gg 1$ only for the thermal SZ and the background offset, as seen previously. For the correlation of the \redmagic samples with stellar density appears more discrepant but is still consistent within $2\sigma$. Similarly, the correlation with \texttt{maglim\_i} is consistent within $1.5\sigma$. We show results for the \redmagic high-luminosity sample, but note that the high-density samples show consistent behavior as well.}
\label{fig:survey_properties_maglim}
\end{figure*}

In Section \ref{sec:sec:sec:sysmaps} we detail the correlations between the number of detected/selected galaxies and the survey property maps; see that Section for more details. An important finding was that the correlations seen in \Balrog were consistent with those measured in the data (except in two cases where we \textit{a priori} expect disagreement due to cosmological correlations). We now repeat this measurement for the two lens sample used in DES Y6: \maglim and \redmagic.

Figure \ref{fig:survey_properties_maglim} repeats the test of Figure \ref{fig:survey_properties} but now using these two lens samples. Note that these samples are a significantly smaller subset of the main Y6 Gold catalog and therefore more sparsely cover the survey footprint. For this reason, we use $\texttt{NSIDE} = 128$ when generating maps for the correlations. In general, we find that the galaxy--survey property correlations are consistent between \Balrog and the data, for both \maglim and \redmagic. The main outliers are the thermal SZ map and the background offset map, where the difference between the correlations --- quantified as a $\chi^2$ metric --- exhibits $\chi^2/N_{\rm dof} \gg 1$. This difference is expected, as we discussed in Section \ref{sec:sec:sec:sysmaps}. We have also repeated this test for the galaxies in each tomographic bin of the \maglim sample (Figure not shown) and find the same qualitative conclusions within each bin. The \maglim sample used in Figure \ref{fig:survey_properties_maglim} has a redshift range of $0.2 < z < 1.05$.

This test is more analogous to the work of \citet{Kong:2024:ObiWan} --- when compared to our analysis of Y6 Gold in Section \ref{fig:survey_properties} --- given their work tested the selection of luminous red galaxies (LRGs), which is a color-based selection. The \maglim sample is primarily a magnitude-based selection, with minor color selections to reduce stellar contamination, while the \redmagic sample is a color-based selection of LRGs. Even after using samples with such selections, we find that the galaxy--survey property correlations are consistent between the real data and the \Balrog synthetic sample. This further corroborates the findings of \citet{Kong:2024:ObiWan}, where the main differences of the galaxy--extinction correlations between the real and synthetic samples were found in the northern Galactic cap (i.e. \textit{not} in the sky region encompassing the DES data).

\section{Photometric performance of the star sample}\label{appx:performancestar}

\begin{figure*}
\centering
\includegraphics[width = 2\columnwidth]{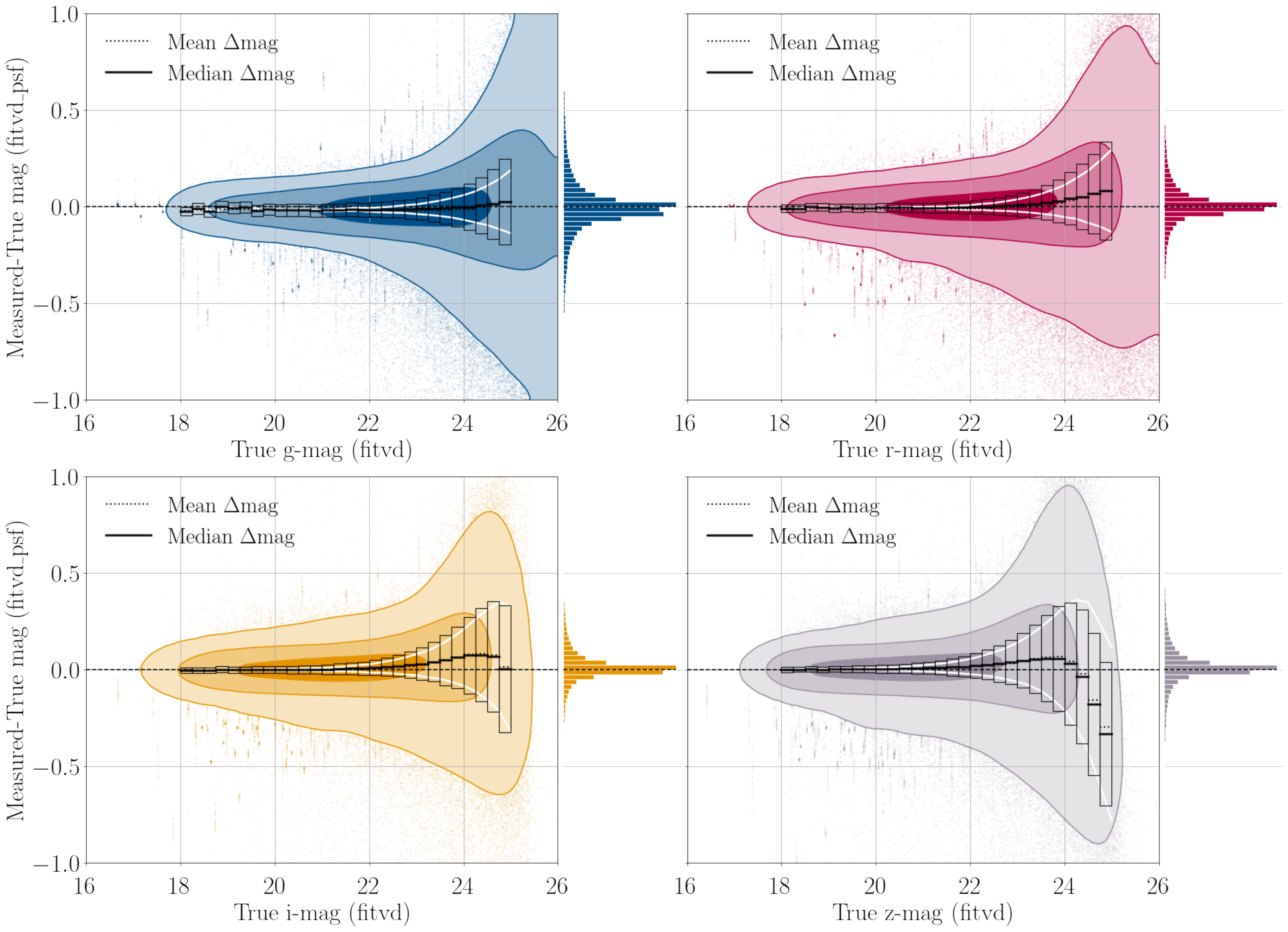}
\caption{Similar to Figure \ref{fig:y6deepmags} but for the star sample. We compute the differences for the \fitvd \textit{psf} magnitudes (instead of the standard model fits) as these are more relevant for stellar studies. The features are all broadly similar to those found in Figure \ref{fig:y6deepmags}, and a detailed numerical comparison is shown in Table \ref{tab:photo_acc}.}
\label{fig:starmags}
\end{figure*}

\begin{figure*}
\centering
\includegraphics[width = 2\columnwidth]{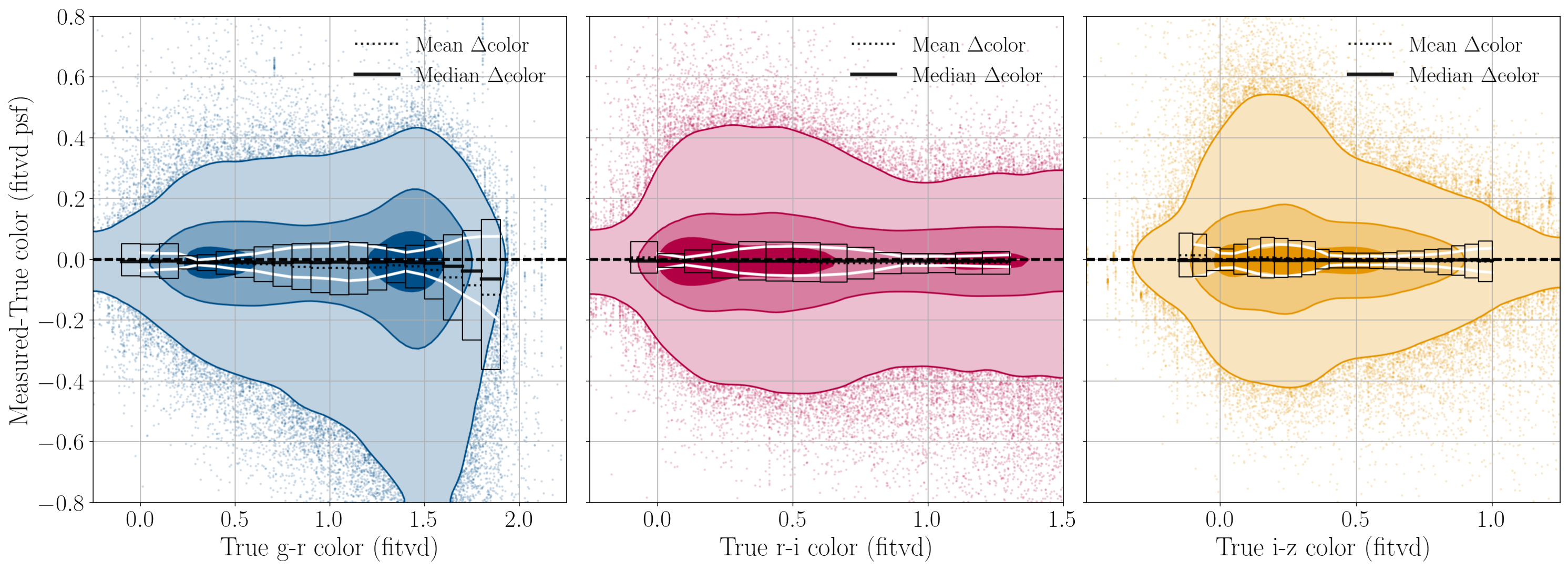}
\caption{Similar to Figure \ref{fig:starmags} but for the \fitvd $g-r$, $r-i$, and $i-z$ colors.}
\label{fig:starcolor}
\end{figure*}

The main body of our work focuses on validating the synthetic galaxy sample, and testing the accuracy of the image processing pipeline in measuring the properties of galaxies. However, a vast variety of science performed with the DES data use star catalogs \citep[\eg][]{Bechtol:2015:Dwarfs, Shipp:Streams:2018, Pieres:2020:MW}.
To aid such science targets, we now explicitly validate the performance of the Y6 processing pipeline in recovering the properties of stars. We stress one notable difference between the validation in our work in comparison to that done in \citetalias{BalrogY3}: the latter performed this test using simple, simulated point-sources as the injected star sample whereas we injected actual measurements of stars from the DES Y3 deep-field catalog \citepalias{DES_DF_Y3}. This deep-field sample is determined using the photometric classifier built in Y3, which provides a $>95\%$ pure star sample down to the 24th magnitude in the $i$-band (see their Figure 15).

Figure \ref{fig:starmags} shows the difference in recovered magnitudes compared to the injected magnitude $\Delta\text{mag}_{\delta}$ as a function of injected magnitude for the \textit{griz} bands. We show the summary statistics of the bias as done in Figure \ref{fig:y6deepmags}. The results are generally similar to that discussed in Figure \ref{fig:y6deepmags}: the pipeline is accurate at the few millimag level for brighter objects, and shows larger biases, at the tens of millimag, for fainter objects. The effects of a detection bias and potentially incorrect background subtraction in \SE, as discussed in Section \ref{sec:sec:performancefid} for the galaxy sample, can explain the slightly negative biases found in this analysis as well. Note that we show the \fitvd PSF magnitudes, which are photometric measurements that use the PSF for the functional form of the light profile and are therefore optimized for studying point sources. We still show the standard \fitvd measurements on the x-axis to facilitate easy comparison with Figure \ref{fig:y6deepmags}.

In comparison with the analysis done on the galaxy sample (section \ref{sec:sec:performancefid}), we find that the deviations in recovered magnitude of the stellar sample has a scatter that is on average half of that measured in the galaxy sample, across the full magnitude range (see Table \ref{tab:photo_acc}). This is expected because the galaxy sample contains injected objects with a wide variety of morphologies, compared to the purer stellar injections we study here, and also because the galaxy samples could be more easily affected by residual light from nearby sources, given they are more extended objects than their stellar counterparts.

In Figure \ref{fig:starcolor}, we plot the difference in measured and input $g-r$, $r-i$, and $i-z$ color as a function of input color. The contours and summary statistics are calculated the same way as with the magnitudes mentioned above (and first discussed in Figure \ref{fig:y6deepmags}), except now we use narrowed magnitude bins, of size 100 millimag for $g-r$ and $r-i$, and 50 millimag for $i-z$. We continue to use only objects with signal-to-noise greater than 5. For the three color definitions we study, the accuracy of the recovered colors is consistently below 10 millimag for most of the color range. For the redder colors (right-end of all panels), the median bias grows to the $5\%$ level for $g-r$ alone. We find no evidence for chromatic effects in the bias as the behavior is consistent across all three color definitions.

\section{Tabulation of photometric accuracy}

Table \ref{tab:photo_acc} presents the photometric accuracy, in the $griz$ bands, as presented in Figure \ref{fig:y6deepmags} and \ref{fig:starmags}. We show the median magnitude difference (in millimag) and scatter on this difference (also in millimag) in the parantheses. The latter is computed by measuring the $68\%$ range and taking the scatter to be half the width of this interval.



\end{document}